\documentclass[12pt]{elsarticle}                                          
\usepackage{axodraw4j}
\usepackage{hyperref}
\usepackage{graphicx}                                                       
\usepackage{a41}                                                             
\usepackage{xcolor}                     
\usepackage{slashed}
\usepackage[rflt]{floatflt}
\usepackage{float}
\usepackage{hyperref}
\hypersetup{colorlinks=true, 
linkcolor=blue, filecolor=blue, urlcolor=mygreen}
\usepackage{breakurl} 
\usepackage{times}       
\usepackage{array}
\usepackage[english]{babel}
\usepackage[latin1]{inputenc}
\usepackage[T1]{fontenc}
\usepackage{ae}
\usepackage{url}
\usepackage{amsmath, amsthm, amssymb}
\usepackage{slashed}

\usepackage{rotating}
\usepackage{graphicx}
\usepackage{comment}
\newcounter{mmacnt}
\def\restartmma{\setcounter{mmacnt}{0}}
\restartmma \catcode`|=\active
\def|#1|{\mathrm{#1}}
\catcode`|=12
\newenvironment{mma}{
\par\smallskip
\catcode`|=\active
\parskip=0pt\parindent=0pt 
\small
\def\In##1\\{%
\def\linebreak{\hfill\break\null\qquad}%
\refstepcounter{mmacnt}
\hangindent=2.5em\hangafter=0
\leavevmode
\llap{\tiny\sffamily In[\arabic{mmacnt}]:=\kern.5em}%
\mathversion{bold}\footnotesize$
\displaystyle##1$\normalsize
\mathversion{normal}\par
 }%
\def\Print##1\\{%
\def\linebreak{\hfill\break}%
\hangindent=2.5em\hangafter=0
\leavevmode ##1\par}%
\def\Out##1\\{%
\def\linebreak{$\hfill\break\null\hfill$}%
\kern\abovedisplayskip\par
\hangindent=2.5em\hangafter=0
\leavevmode
\llap{\tiny\sffamily Out[\arabic{mmacnt}]=\kern.5em}
\footnotesize$\displaystyle##1$
\normalsize\hfill\null\par
\kern\belowdisplayskip
}%
\def\Warning##1##2\\{%
\def\linebreak{\hfill\break}%
\hangindent=2.5em\hangafter=0
\leavevmode
{\scriptsize##1 : ##2}\par}%
}{%
\par\smallskip
}

\usepackage{color}
\newenvironment{fshaded}{%
\MakeFramed {\FrameRestore}
}%
{\endMakeFramed}

\newcommand{\n}{\nonumber}
\makeatletter
\def\ps@pprintTitle{%
\let\@oddhead\@empty
\let\@evenhead\@empty
\def\@oddfoot{\reset@font\hfil\thepage\hfil}
\let\@evenfoot\@oddfoot
}
\makeatother
\begin{document}   
\begin{frontmatter}
\title{\textbf{General one-loop formulas 
for $H\rightarrow f\bar{f}\gamma$ and its 
applications}}
\author[0]{Vo Van On}
\ead{onvv@tdmu.edu.vn}
\author[4]{Dzung Tri Tran}
\author[4]{Chi Linh Nguyen}
\author[1,2]{Khiem Hong Phan}
\ead{phanhongkhiem@duytan.edu.vn}
\address[0]{\it Institute of Applied Technology, 
Thu Dau Mot University, Thu Dau Mot City, 
Binh Duong Province 75000, Vietnam}
\address[4]{\it University of Science Ho Chi Minh City, $227$ 
Nguyen Van Cu, District $5$, Ho Chi Minh City, Vietnam}
\address[1]{\it Institute of Fundamental and Applied Sciences, 
Duy Tan University, Ho Chi Minh City $700000$, Vietnam}
\address[2]{Faculty of Natural Sciences, Duy Tan University, 
Da Nang City $550000$, Vietnam}
\pagestyle{myheadings}
\markright{}
\begin{abstract} 
We present general one-loop contributions to the decay processes 
$H\rightarrow f\bar{f}\gamma$ including all possible the exchange 
of the additional heavy vector gauge bosons, heavy fermions, 
and charged (also neutral) scalar particles in the loop 
diagrams. As a result, the analytic results 
are valid in a wide class of beyond the standard models. Analytic 
formulas for the form factors are expressed in terms of
Passarino-Veltman functions in the standard
notations of {\tt LoopTools}. Hence, the decay rates can be computed 
numerically by using this package. The computations are then applied 
to the cases of the standard model, $U(1)_{B-L}$
extension of the standard model as well as two Higgs doublet model. Phenomenological results of the decay processes for all
the above models are studied. We observe that the effects of new 
physics are sizable contributions and these can be probed at 
future colliders. 
\end{abstract}
\begin{keyword} 
{\footnotesize \it
Higgs phenomenology, One-loop Feynman integrals,  
Analytic methods for Quantum Field Theory, 
Dimensional regularization.
}
\end{keyword}
\end{frontmatter}
\section{Introduction} 
After discovering the standard model-like (SM-like) 
Higgs boson~\cite{ATLAS:2012yve,CMS:2012qbp}, 
one of the main purposes at future colliders
like the high luminosity large hadron 
collider (HL-LHC)~\cite{Liss:2013hbb,CMS:2013xfa} as well as 
future lepton colliders~\cite{Baer:2013cma} 
is to probe the properties of this boson 
(mass, couplings, 
spin and parity, etc). In the experimental programs, 
the Higgs productions and its decay rates should be 
measured as precise as possible. Based on the 
measurements, we can verify the nature of the 
Higgs sector. In other words we can understand 
deeply the dynamic of the 
electroweak symmetry breaking. It is well-known that 
the Higgs 
sector is selected as the simplest case in the 
standard model (SM) which there is only a scalar 
doublet field. From theoretical viewpoints, there are no 
reasons for this simplest choice. Many of beyond
the standard models (BSMs) have extended the Higgs sector 
(some of them have also expanded gauge sectors, 
introduced mass terms of neutrinos, etc). In these 
models many new particles are proposed, for examples, 
new heavy gauge bosons, charged and neutral scalar Higgs 
as well as new heavy fermions. These new particles 
may also contribute to the productions and decay 
of Higgs boson. It means that the more precise data 
and new theoretical approaches on the Higgs productions 
and its decay rates could provide us a crucial tool to 
answer the nature of the Higgs sector and, more important, 
to extract the new physics contributions.

Among all the Higgs decay channels, the processes 
$H\rightarrow f\bar{f}\gamma$ are great of interest
at the colliders by following reasons. Firstly, 
the decay channels can be measured at the 
large hadron collider~\cite{CMS:2015tzs,
CMS:2017dyb, CMS:2018myz,ATLAS:2021wwb}. 
Therefore, the processes can be used to test 
the SM at the high energy regions. 
Secondly, many of new particles 
as mentioned in the beginning of this section
may propagate in the loop diagrams of the decay 
processes. Subsequently, the decay rates could 
provide a useful tool for constraining new physic
parameters. Last but not least, 
apart from the SM-like Higgs boson, new neutral Higgs 
bosons in BSMs may be mixed with the SM-like 
one. These effects can be also observed
directly by measuring of the decay rates
of $H\rightarrow f\bar{f}\gamma$. As above reasons, 
the detailed theoretical evaluations for one-loop 
contributions to the decay of Higgs to fermion 
pairs and a photon within the SM and its 
extensions are necessary.

Theoretical implications for the decay 
$H\rightarrow f\bar{f}\gamma$ in the SM at the LHC 
have studied in Refs.~\cite{Chen:2012ju,Gainer:2011aa,
Korchin:2014kha}. Moreover, there have been many 
available computations for one-loop contributing to 
the decay processes $H\rightarrow f\bar{f}\gamma$ 
within the SM framework~\cite{Abbasabadi:1995rc,
Djouadi:1996ws,Abbasabadi:2000pb,Dicus:2013ycd,
Sun:2013rqa,Passarino:2013nka,Dicus:2013lta,Kachanovich:2020xyg}. 
The same evaluations for the Higgs productions at $e\gamma$ 
colliders have proposed in~\cite{Watanabe:2013ria,Watanabe:2014xaa}. 
While one-loop corrections to $H\rightarrow f\bar{f}\gamma$
in the context of the minimal super-symmetric standard model 
Higgs sector have computed in~\cite{Li:1998rp}.
Furthermore, one-loop contributions
for CP-odd Higgs boson productions in $e\gamma$ 
collisions have carried out in~\cite{Sasaki:2017fvk}. 
In this article, we present
general formulas for one-loop contributing to the 
decay processes $H\rightarrow f\bar{f}\gamma$.  
The analytic results presented in the current paper are 
not only valid in the SM but also in many of BSMs 
in which new particles are proposed such as 
heavy vector bosons, heavy fermions, and charged (neutral) 
scalar particles that may propagate in the loop diagrams 
of the decay processes. The analytic 
formulas for the form factors are expressed in terms
of Passarino-Veltman (PV) functions in standard notations of 
{\tt LoopTools}~\cite{Hahn:1998yk}. As a result, 
they can be evaluated numerically by using this package. 
The calculations are then applied to the SM and many of 
beyond the SM such as the $U(1)_{B-L}$ extension of 
the SM~\cite{Basso:2008iv}, two Higgs doublet models 
(THDM)~\cite{Branco:2011iw}. 
Phenomenological results of the decay processes for 
these models are also studied.

We also stress that our analytical results in the present 
paper can be also applied for many of BSMs framework. 
In particular, in the super-symmetry models, many 
super-partners of fermions and gauge bosons are introduced.
Furthermore, with extending the Higgs sector, we encounter
charged and neutral Higgs bosons in this framework. 
There are exist the extra charged gauge bosons in many 
electroweak gauge extensions, for examples, 
the left-right models (LR) constructed from the 
$SU(2)_L\times SU(2)_R\times U(1)_Y$~\cite{Pati:1974yy,
Mohapatra:1974gc, Senjanovic:1975rk}, the 3-3-1 models 
($SU(3)_L\times U(1)_X$)~\cite{Singer:1980sw, Valle:1983dk,
Pisano:1991ee, Frampton:1992wt, Diaz:2004fs,Fonseca:2016tbn, Foot:1994ym}, 
the $3$-$4$-$1$ models ($SU(4)_L\times U(1)_X$)~\cite{Foot:1994ym,
Sanchez:2004uf, Ponce:2006vw, Riazuddin:2008yx, Jaramillo:2011qu,
Long:2016lmj}, etc. Analytic results in this paper 
already include the contributions these mentioned particles which 
may also exchange in the loop diagrams of the 
aforementioned decay processes. Phenomenological results for 
the decay processes in the above models are great of interest
in future. These topic will be devoted in our future publications.

The layout of the paper is as follows: We first write down the general 
Lagrangian and introduce the notation for the calculations in the section $2$. 
We then present the detailed calculations for one-loop 
contributions to $H\rightarrow f\bar{f}\gamma$ in section $3$. 
The applications of this work to the SM, $U(1)_{B-L}$ extensions 
of the SM and THDM are also studied in this section. 
Phenomenological results for these models are analysed at the 
end of section $3$. Conclusions and outlook are devoted in the 
section $4$. In appendices, the checks for the computations are shown. 
We then review briefly $U(1)_{B-L}$ extensions of the SM and THDM in 
the appendices. Finally, Feynman rules and all involving 
couplings in the decay processes are shown. 
\section{Lagrangian and notations}     
In order to write down the general form of Lagrangian for 
a wide class of the BSMs, we start from the well-known 
contributions appear in the SM. We then consider the 
extra forms that extended from the SM. For example, the 
two Higgs doublet model~\cite{Branco:2011iw} 
adding a new Higgs doublet that predict new charged 
and neutral scalar Higgs bosons; 
a model with a gauge symmetry $U(1)_{B-L}$ which proposes
a neutral gauge boson $Z'$~\cite{Basso:2008iv,Michaels:2020fzj}, 
neutral Higgs; a minimal left-right models with a new non-Abelian 
gauge symmetry for electroweak interactions 
$SU(2)_L\times SU(2)_R\times U(1)_{B-L}$~\cite{Pati:1974yy,
Mohapatra:1974gc, Senjanovic:1975rk} introducing many 
new particles including charged gauge bosons, neutral gauge 
bosons and charged Higgs bosons. The mentioned particles 
give one-loop contributions to the decays under consideration.

In this section, Feynman  
rules for the decay channels $H\rightarrow f\bar{f}\gamma$ 
are derived for the most general extension of the SM with
considering all possible contributions from the mentioned 
particles. In this computation, we denote $V_i, V_j$ 
for extra charged gauge bosons, $V_k^0$ for neutral gauge 
bosons. Moreover, $S_i, S_j\; (S_k^0)$ are charged (neutral) 
Higgs bosons respectively and $f_i, f_j$ show for fermions. 
In general, the classical Lagrangian contains the 
following parts:
\begin{eqnarray}
	\mathcal{L} = \mathcal{L}_{f} 
	+ \mathcal{L}_{G} +\mathcal{L}_{\Phi} 
	+\mathcal{L}_{Y}. 
\end{eqnarray}
Where the fermion sector is given
\begin{eqnarray}
	\mathcal{L}_{f}  = \bar{\psi}_f i \slashed D \psi_f
\end{eqnarray}
with $D_{\mu} = \partial_{\mu} - i g T^a V_{\mu}^a + \cdots$. 
In this formula, $T^a$ is generator of the corresponding
gauge symmetry. The gauge sector reads
\begin{eqnarray}
	\mathcal{L}_G = -\frac{1}{4}\sum\limits_{a} V^a_{\mu\nu}V^{a, \mu\nu} 
	+ \cdots 
\end{eqnarray}
where $V^a_{\mu\nu} = \partial_{\mu}V^a_{\nu} -\partial_{\nu}V^a_{\mu}+
gf^{abc}V^b_{\mu}V^c_{\mu}$ with $f^{abc}$ is structure constant of 
the corresponding gauge group. The scalar sector is expressed
as follows:
\begin{eqnarray}
\mathcal{L}_{\Phi} = \sum\limits_{\Phi} Tr[\left(D_{\mu}\Phi
\right)^{\dag}\left(D^{\mu}\Phi\right) ] - V(\Phi). 
\end{eqnarray}
We then derive all the couplings from the full Lagrangian. 
The couplings are parameterized in general forms and 
presented by following:
\begin{itemize}
	\item By expanding the fermion sector, we can derive the vertices 
	of vector boson $V$ with fermions. In detail, the interaction terms
	are parameterized as
	\begin{eqnarray}
		\mathcal{L}_{Vff} = \sum\limits_{f_i,f_j,V} \bar{f}_i
		\gamma^{\mu}( g_{Vff}^L P_L +  g_{Vff}^R P_R)f_j V_{\mu}+\cdots
	\end{eqnarray}
	with $P_{L,R} = (1\mp \gamma_5)/2$. 
	\item Trilinear gauge and quartic gauge couplings are expanded from 
	the gauge sector:
	\begin{eqnarray}
		\mathcal{L}_{VVV, VVVV} &=& \sum\limits_{V_k^0,V_i,V_j} 
		g_{V_k^0V_iV_j}\Big[\partial_{\mu} V^0_{k,\nu} V_{i}^{\mu}V_j^{\nu} 
		+ V_{k,\nu}^0 V_{i}^{\mu}\partial^{\nu}V_{j,\mu} + \cdots \Big]
		\n \\
		&& +\sum\limits_{V_k^0,V_l^0,V_i,V_j} 
		g_{V_k^0V_l^0V_iV_j}\Big[V^0_{k,\mu} V_{l,\nu}^0 V_{i}^{\mu}V_j^{\nu}
		+ \cdots \Big] + \cdots
	\end{eqnarray}
	\item Next, the couplings of scalar $S$ to fermions are taken from 
	the Yukawa part $\mathcal{L}_Y$. The interaction terms are expressed 
	as follows:
	\begin{eqnarray}
		\mathcal{L}_{Sf_if_j} = \sum\limits_{f_i,f_j,S} \bar{f}_i
		(g_{Sff}^L P_L +  g_{Sff}^R P_R)f_j S + \cdots 
	\end{eqnarray}
	\item The couplings of scalar $S$ to vector boson $V$ can be derived
	from the kinematic term of the Higgs sector. In detail, we have 
	the interaction terms 
	\begin{eqnarray}
		\mathcal{L}_{SVV, SSV, SSVV} 
		&=& \sum\limits_{S,V_i, V_j} g_{SV_iV_j} SV_{i}^{\mu}V_{j, \mu}
		+ \sum\limits_{S_i, S_j, V} g_{S_iS_jV}[(\partial_{\mu}S_i) S_j
		- (\partial_{\mu}S_j) S_i] V^{\mu}
		\n\\
		&& 
		+ \sum\limits_{S_i,S_j,V_k, V_l} g_{S_iS_jV_kV_l} S_iS_jV_{k}^{\mu}V_{l, \mu}+\cdots 
	\end{eqnarray}
	\item Finally, the trilinear scalar and quartic scalar interactions are from 
	the Higgs potential $V(\Phi)$. The interaction terms are written as 
	\begin{eqnarray}
		\mathcal{L}_{SSS, SSSS} 
		&=& \sum\limits_{S_i,S_j, S_k} g_{S_iS_jS_k} S_iS_{j}S_k +
		\sum\limits_{S_i,S_j, S_k, S_l} g_{S_iS_jS_k,S_l} S_iS_{j}S_kS_l 
		+ \cdots
	\end{eqnarray}
\end{itemize}
All of the Feynman rules corresponding to the above 
couplings giving one-loop contributions to the 
SM-like Higgs decays $H\to f\bar{f} \gamma$
are collected in appendix $D$. In detail,
the propagators involving the decay processes 
in the unitary gauge
are shown in Table~\ref{Feynman rules table}.
All the related couplings involving the decay
channels are parameterized in general forms
which are presented in Table~\ref{couplings table} 
(we refer appendices $B$ and $C$ for two 
typical models). 
\section{Calculations}   
\begin{figure}[]
\centering
\includegraphics[scale=0.32]
{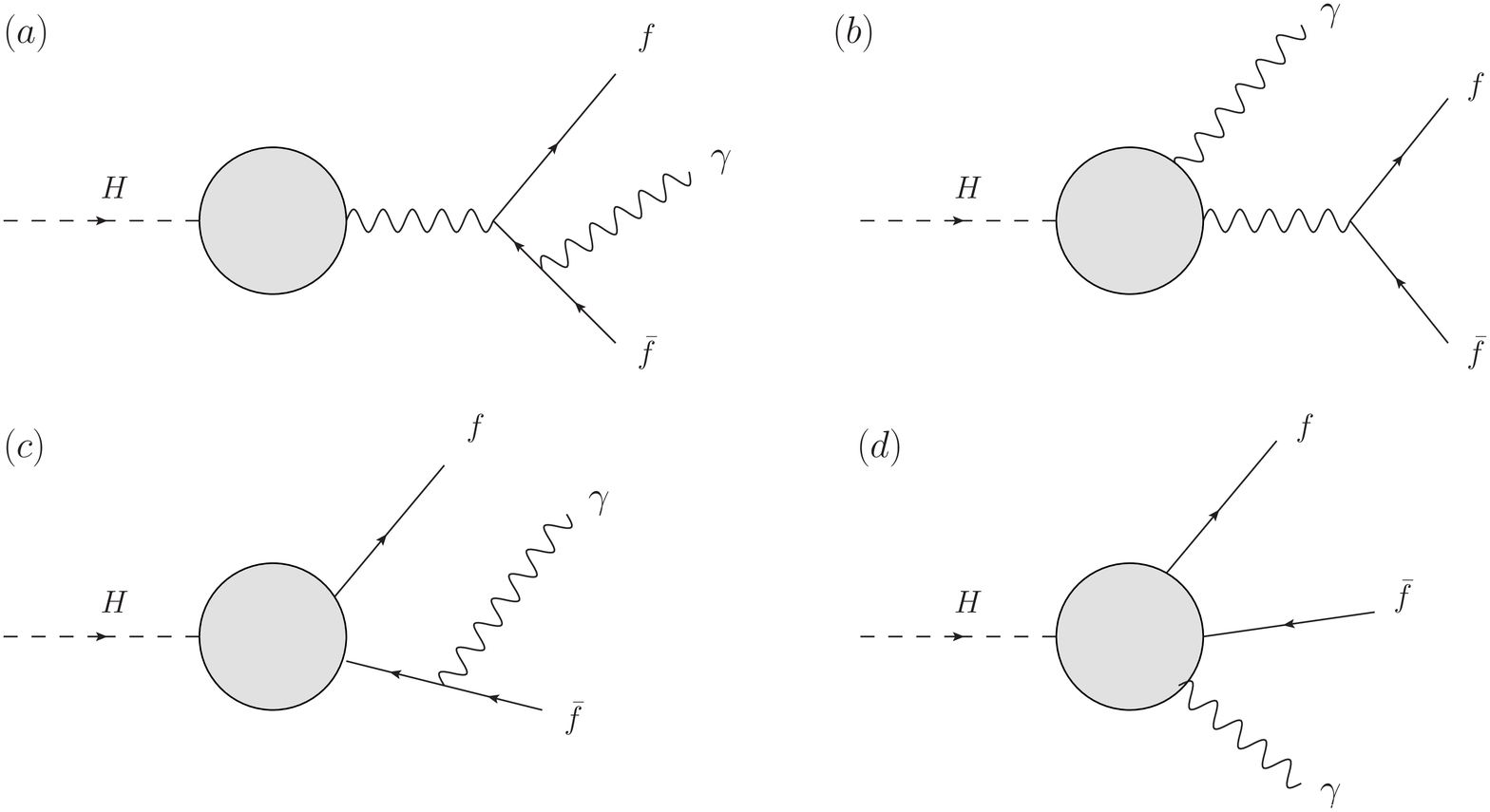} \\
\includegraphics[scale=0.32]
{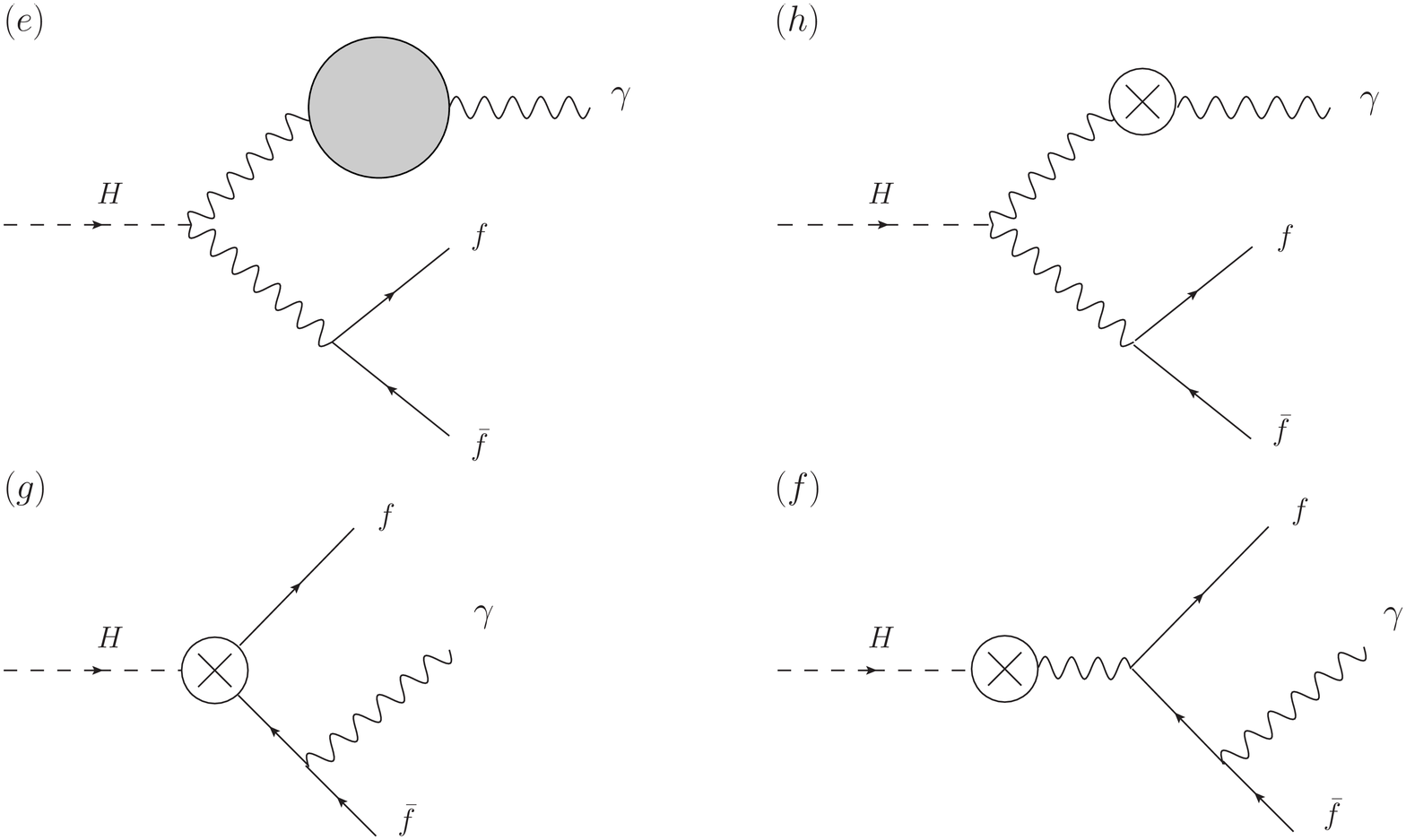} \\
\includegraphics[scale=0.35]
{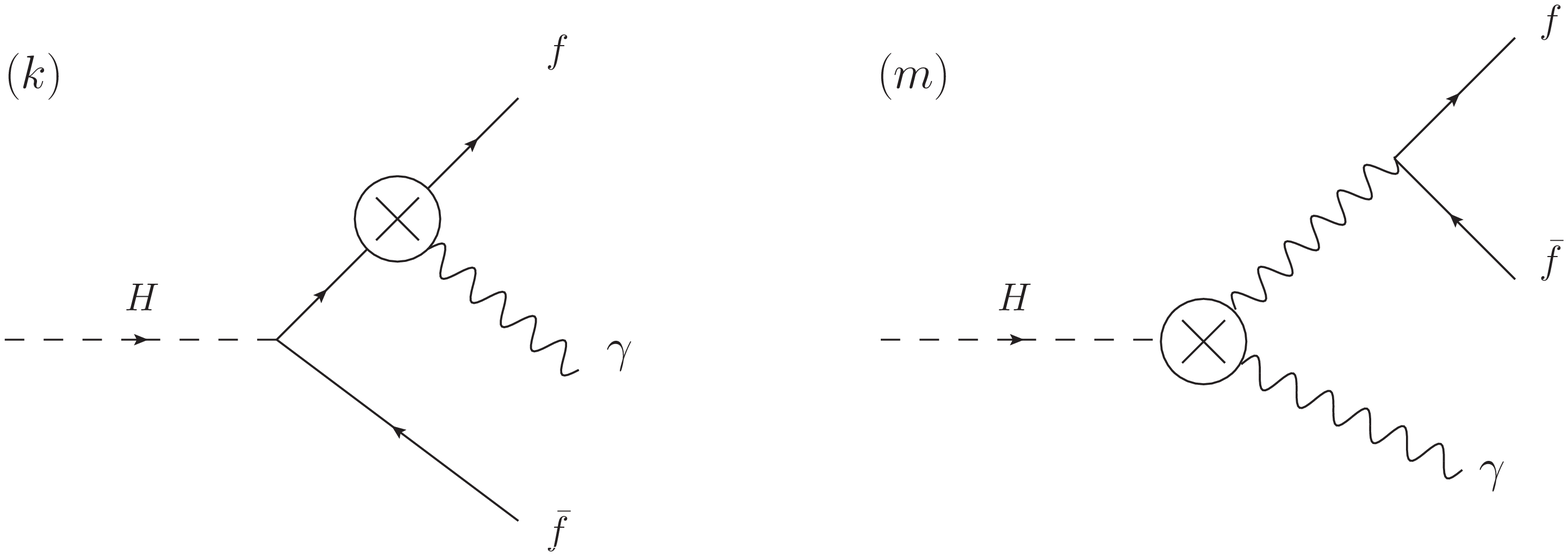}
\caption{\label{feynFFg} Types of Feynman 
diagrams involving to the SM-like Higgs decays 
$H\to f\bar{f} \gamma$}
\end{figure}

In this section, one-loop contributions to the 
decay processes $H\rightarrow f(q_1)\bar{f}(q_2)\gamma (q_3)$
are calculated in detail. In the present paper,
we consider the computations in the limit of 
$m_f\rightarrow 0$. All Feynman diagrams involving 
these processes can be grouped in to the following 
classes (seen Fig.~\ref{feynFFg}). 
For on-shell photon, we confirm that the 
contributions of diagrams $(e+h)$ will be 
vanished. One can neglect the Yukawa coupling 
$y_f$ (since $m_f \rightarrow 0$) in this computation. 
As a result, the contributions of diagrams $(g+k+m)$ can 
be omitted. Furthermore, the diagrams $(a+f)$ are not 
contributed to the amplitude. Hence, we only have the 
contributions of $(b+c+d)$ which are separated into 
two kinds of the contributions. The first one is to 
the topology $b$ which is called $V_{k}^{0*}$ pole 
contributions. The second type (diagrams $c$ and $d$) 
belongs to the non-$V_{k}^{0*}$ pole 
contributions. We remind that $V_{k}^{0*}$ can include 
both $Z, \gamma$ in the SM and the arbitrary neutral vector 
boson $Z'$ in many of the BSMs. General one-loop amplitude 
which obeys the invariant Lorentz structure
can be decomposed as follows~\cite{Kachanovich:2020xyg}:
\begin{eqnarray}
 \mathcal{A}_{\text{loop}} &=&\sum\limits_{k=1}^2 
 \Big\{
[q_3^{\mu}q_k^{\nu} - g^{\mu\nu}q_3\cdot q_k] \bar{u}(q_1)
(F_{k,R} \gamma_{\mu}P_R + F_{k,L} \gamma_{\mu}P_L) v(q_2) 
\Big\}\varepsilon^{*}_{\nu}(q_3).
\end{eqnarray}
In this equation, all form factors are computed as follows:
\begin{eqnarray}
\label{FLR}
 F_{k,L/R} &=& F_{k,L/R}^{\text{$V_{k}^{0*}$-poles}}
 +  F_{k,L/R}^{\text{Non-$V_{k}^{0*}$} }
\end{eqnarray}
for $k=1,2$. 
Kinematic invariant variables 
involved in the decay processes are included: 
$q^2 = q_{12} = (q_1+q_2)^2$, $q_{13} = (q_1+q_3)^2$
and $q_{23} = (q_2+q_3)^2$.

We first write down all Feynman amplitudes for the above 
diagrams. With the help of {\tt Package-X}~\cite{Patel:2015tea}, 
all Dirac traces and Lorentz contractions in $d$ dimensions
are handled. Subsequently, the amplitudes are then 
written in terms of tensor one-loop integrals. 
By following tensor reduction for 
one-loop integrals in~\cite{Denner:2005nn} 
(the relevant tensor reduction formulas 
are shown in appendix $A$), 
all tensor one-loop integrals are expressed in terms  
of PV-functions. 
\subsection{$V_{k}^{0*}$ pole contributions} 
In this subsection, we first arrive at 
the $V_{k}^{0*}$ pole contributions which are 
corresponding to the diagram $b$. In this group 
of Feynman diagrams, it is easily to confirm 
that the form factors follow 
the below relation:  
\begin{eqnarray}
F_{k,L/R}^{\text{$V_{k}^{0*}$-poles}}
=F_{1,L/R}^{\text{$V_{k}^{0*}$-poles}}
=F_{2,L/R}^{\text{$V_{k}^{0*}$-poles}} 
\end{eqnarray}
Their analytic results are will be shown 
in the following subsections. All possible
particles exchanging in the loop diagrams
are included. We emphasize that analytic
expressions
for the form factors presented in this 
subsection cover the results in 
Ref.~\cite{Phan:2021xwc}.
It means that we can reduce to the results 
for $H\to \nu_l\bar{\nu}_{l} \gamma$
of Ref.~\cite{Phan:2021xwc} 
by setting $f$ to $\nu_l$
and replacing the corresponding couplings. 
Furthermore, all analytic formulas 
shown in the following subsection cover 
all cases of $V_{k}^{0*}$ poles. For instance,
when $V_{k}^{0*} \rightarrow \gamma^*$, we then set 
$M_{V^0_k}=0$, $\Gamma_{V^0_k}=0$. In addition, 
$V_{k}^{0*}$ becomes $Z$ (or $Z'$) boson, 
we should fix 
$M_{V^0_k}=M_Z$ and 
$\Gamma_{V^0_k}=\Gamma_{Z}$ (or $Z'$) respectively.
\begin{figure}
\centering
\includegraphics[width=12cm, height=7cm]
{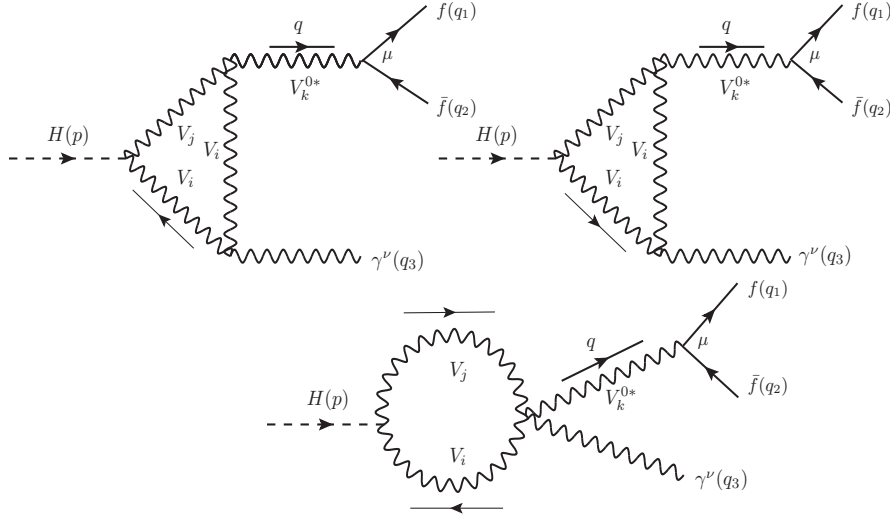}
\caption{\label{ViVj-dig} One-loop triangle 
diagrams with exchanging
vector boson $V_{i,j}$ particles in the loop.}
\end{figure}
\begin{figure}
\centering
\includegraphics[width=12cm, height=7cm]
{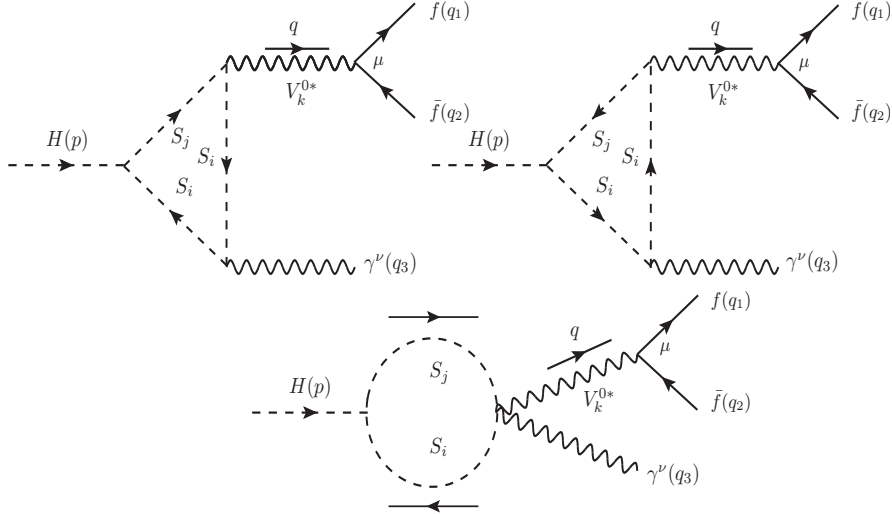}
\caption{\label{SiSj-dig} One-loop triangle 
diagrams with exchanging
scalar boson $S_{i,j}$ particles in loop.}
\end{figure}
\begin{figure}
\centering
\includegraphics[width=12cm, height=4cm]
{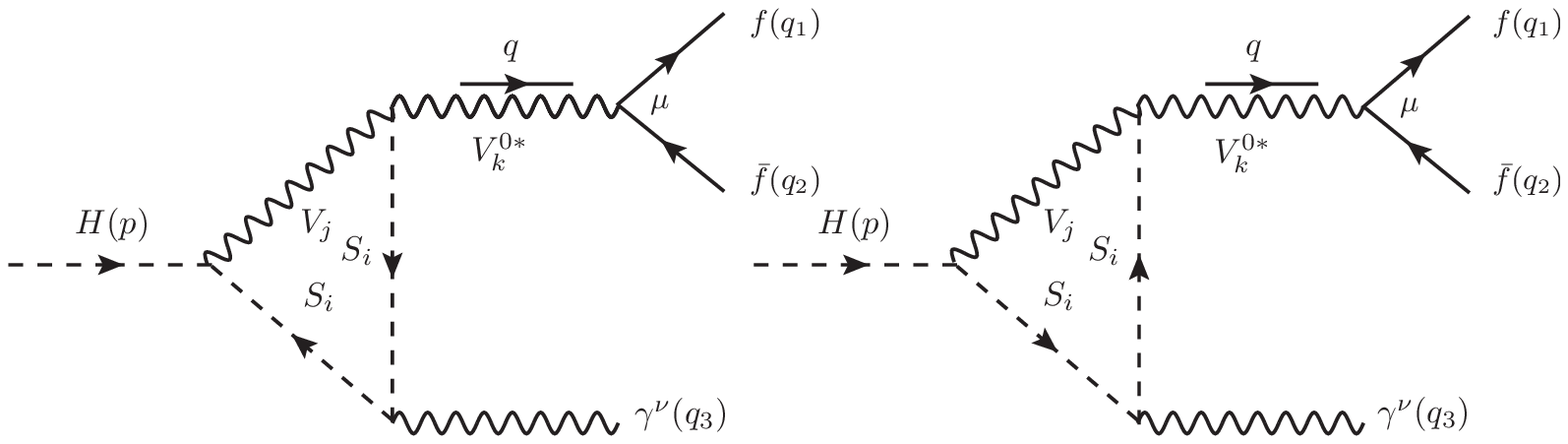}
\caption{\label{SSiVj-dig} One-loop triangle 
diagrams with exchanging two scalar bosons 
$S_{i}$ and a vector boson $V_j$ particles 
in the loop.}
\end{figure}
\begin{figure}[]
\centering
\includegraphics[width=10cm, height=4cm]
{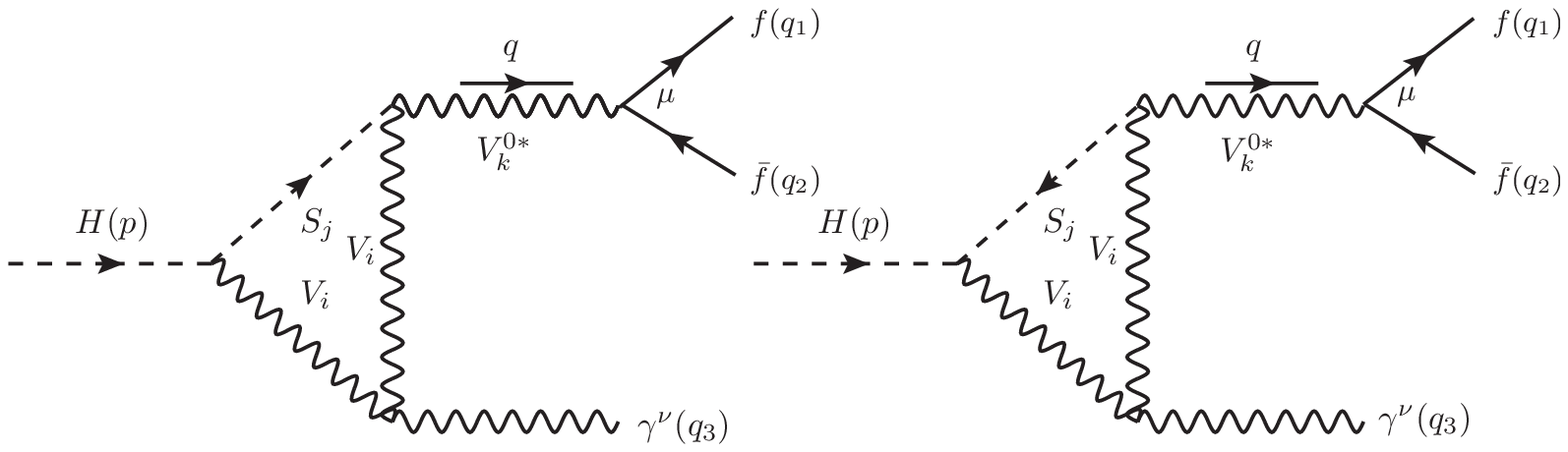}
\caption{\label{SiVVj-dig} One-loop triangle 
diagrams with exchanging a scalar boson $S_{j}$ 
and two vector bosons $V_i$ in the loop.}
\end{figure}
\begin{figure}[]
\centering
\includegraphics[width=10cm, height=4cm]
{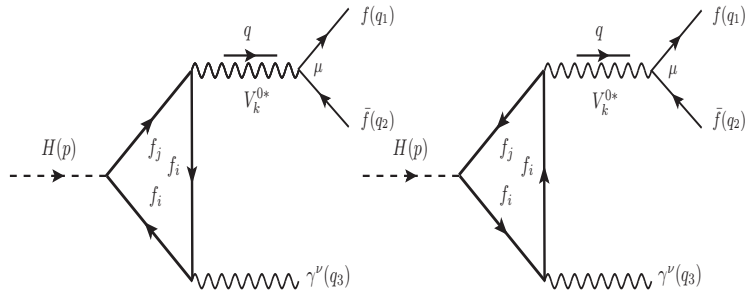}
\caption{\label{fifj-dig} One-loop 
triangle diagrams with exchanging
charged fermion $f_{i,j}$ particles in loop.}
\end{figure}

We begin with 
one-loop triangle Feynman diagrams which 
all vector bosons $V_{i,j}$ are in the loop 
(seen Fig.~\ref{ViVj-dig}). 
One-loop form factors of this group of Feynman 
diagrams are expressed in terms of the PV 
functions as follows:
\begin{eqnarray}
\label{ViVj-poles}
F_{k,L}^{\text{$V_{k}^{0*}$-poles}}|_{V_i,V_j} 
&=&\dfrac{eQ_V}{16\pi^2} 
\sum\limits_{V_i,V_j, V^0_{k}}
\dfrac{g_{HViVj} \; 
g_{V^0_k f f}^{L} }{M_{V_i}^2 M_{V_j}^2
(q^2 - M_{V^0_k}^2 + i\Gamma_{V^0_k}M_{V^0_k} )}
\times
\n \\
&&
\hspace{-1.5cm} 
\times
\Bigg\{
\Big[
2 g_{V^0_k A V_i V_j} (M_H^2 - M_{V_j}^2)
+g_{V^0_k V_i V_j} (M_H^2
+ M_{V_i}^2
+ M_{V_j}^2)
\Big] 
B_{11}(M_H^2,M_{V_i}^2,M_{V_j}^2)
\n \\
&&\hspace{-1.cm}
+\Big[
g_{V^0_k V_i V_j} (M_H^2
+3 M_{V_i}^2
- M_{V_j}^2)
-2 g_{V^0_k A V_i V_j} M_{V_i}^2
\Big]
 B_1(M_H^2,M_{V_i}^2,M_{V_j}^2)
\n \\
&&\hspace{-1.cm} 
+ 2 M_{V_i}^2
(g_{V^0_k V_i V_j} - g_{V^0_k A V_i V_j}) 
B_0(M_H^2,M_{V_i}^2,M_{V_j}^2)
\n \\
&&\hspace{-1.cm} 
+2 g_{V^0_k A V_i V_j}  
\Big[
M_H^2 B_{111}
+ B_{00}
+2 B_{001}
\Big] (M_H^2,M_{V_i}^2,M_{V_j}^2)
\\
&&\hspace{-1.cm} 
+4 g_{V^0_k V_i V_j} M_{V_i}^2 \Big(
M_{V_i}^2+3 M_{V_j}^2- q^2
\Big)
 C_0(0,q^2,M_H^2,M_{V_i}^2,M_{V_i}^2,M_{V_j}^2)
\n \\
&&\hspace{-1.cm} 
+2 g_{V^0_k V_i V_j}
\Big[ 
M_H^2 (M_{V_i}^2
+ M_{V_j}^2
- q^2)
+ M_{V_i}^4
+ M_{V_j}^4
+(4 d -6) M_{V_i}^2 M_{V_j}^2
\n \\
&&\hspace{2cm}
- q^2 (M_{V_i}^2 
+ M_{V_j}^2)
\Big]
(C_{22}+C_{12})
(0,q^2,M_H^2,M_{V_i}^2,M_{V_i}^2,M_{V_j}^2)
\n \\
&&\hspace{-1.cm} 
+2 g_{V^0_k V_i V_j} \Big[
M_H^2 (M_{V_i}^2
+ M_{V_j}^2
- q^2)
+3 M_{V_i}^4
- M_{V_j}^4
+(4 d -6) M_{V_i}^2 M_{V_j}^2
\n \\
&&\hspace{3cm}
- q^2 (3 M_{V_i}^2 
- M_{V_j}^2 )
\Big] 
C_2(0,q^2,M_H^2,M_{V_i}^2,M_{V_i}^2,M_{V_j}^2)
\Bigg\},
\n\\
\n \\
F_{k,R}^{\text{$V_{k}^{0*}$-poles}}|_{V_i,V_j}
&=&F_{k,L}^{\text{$V_{k}^{0*}$-poles}}|_{V_i,V_j}
\Big( g_{V^0_k ff}^{L} \rightarrow
g_{V^0_k ff}^{R}\Big). 
\end{eqnarray}
The results are written in terms of $B$- 
and $C$-functions. We note that one-loop
amplitude for each diagram in 
Fig.~\ref{ViVj-dig} may decompose
into tensor one-loop up to rank $R=6$. 
However, after taking into account 
all diagrams, the amplitude for this 
subset Feynman diagrams is only expressed 
in terms of the tensor integrals up to 
rank $R=2$. As a result, we have up to
$C_{22}$-functions contributing to 
the form factors. Furthermore, some of 
them may contain UV divergent but after 
summing all these functions, the final 
results are finite. The topic will be 
discussed at the end of this section.

We next concern one-loop triangle Feynman 
diagrams with $S_i,\; S_j$ in the loop 
(as described in Fig.~\ref{SiSj-dig}). 
The corresponding one-loop 
form factors are given:
\begin{eqnarray}
F_{k,L}^{\text{$V_{k}^{0*}$-poles}}|_{S_i,S_j} 
&=&\dfrac{eQ_S}{4\pi^2}\sum\limits_{S_i,S_j, V^0_{k}}
\dfrac{g_{HS_iS_j} \, g_{V^0_k S_iS_j} \, 
g_{V^0_k ff}^L }{M_H^2 q^2 (M_H^2-q^2)
(q^2 - M_{V^0_k}^2 + i\Gamma_{V^0_k}M_{V^0_k} )}
\times
\\
&&\hspace{0.cm} \times
\Bigg\{
(q^2-M_H^2) \Big[A_0(M_{S_i}^2) - A_0(M_{S_j}^2) \Big]
\nonumber \\
&&\hspace{0.5cm} 
+
M_H^2 (M_{S_i}^2-M_{S_j}^2-q^2) B_0(q^2,M_{S_i}^2,M_{S_j}^2)
\nonumber\\
&&\hspace{0.5cm}
+q^2 (M_H^2-M_{S_i}^2+M_{S_j}^2) B_0(M_H^2,M_{S_i}^2,M_{S_j}^2)
\n \\
&&\hspace{0.5cm}
+2 M_H^2 q^2 (M_H^2-q^2) 
C_{12}(0,q^2,M_H^2,M_{S_i}^2,M_{S_i}^2,M_{S_j}^2)
\Bigg\},
\n \\
\n \\
F_{k,R}^{\text{$V_{k}^{0*}$-poles}}|_{S_i,S_j}
&=&F_{k,L}^{\text{$V_{k}^{0*}$-poles}}|_{S_i,S_j}
\Big( g_{V^0_k ff}^{L} \rightarrow
g_{V^0_k ff}^{R}\Big). 
\end{eqnarray}
Similarly, we have the contributions of one-loop 
triangle Feynman diagrams with exchanging scalar 
boson $S_{i}$ and vector boson $V_{j}$ in the loop. 
The Feynman diagrams are depicted as in 
Fig.~\ref{SSiVj-dig}. Applying the same 
procedure, one has the form factors
\begin{eqnarray}
F_{k,L}^{\text{$V_{k}^{0*}$-poles}}|_{S_i,V_j} 
&=&
\dfrac{eQ_V}{8 \pi ^2 }
\sum\limits_{S_i,V_j, V^0_{k}}
\dfrac{g_{HS_iV_j} \, g_{V^0_kS_iV_j} \, 
g_{V^0_kff}^L}{M_H^2 M_{V_j}^2 q^2 (q^2-M_H^2)
(q^2 - M_{V^0_k}^2 + i\Gamma_{V^0_k}M_{V^0_k} )}
\times
 \\
&&\hspace{0.0cm} \times
\Bigg\{
(q^2-M_H^2) (M_H^2-M_{S_i}^2+M_{V_j}^2) 
\Big[A_0(M_{S_i}^2) - A_0(M_{V_j}^2)\Big]
\nonumber\\
&&\hspace{0.0cm}
+
M_H^2 \Big[
q^2 (M_{S_i}^2+3 M_{V_j}^2-M_H^2)
\nonumber\\
&&\hspace{1.3cm}
+(M_{S_i}^2-M_{V_j}^2) (M_H^2-M_{S_i}^2+M_{V_j}^2)
\Big] 
B_0(q^2,M_{S_i}^2,M_{V_j}^2)
\n \\
&&\hspace{0.0cm}
+q^2 
\Big[M_H^2- (M_{S_i}+M_{V_j})^2 \Big]
\nonumber\\
&&\hspace{2.0cm}\times
\Big[M_H^2-(M_{S_i}-M_{V_j})^2 \Big]
B_0(M_H^2,M_{S_i}^2,M_{V_j}^2)
\n \\
&&\hspace{0cm}
+2 M_H^2 q^2 (M_H^2-q^2) (M_H^2-M_{S_i}^2+M_{V_j}^2) 
\nonumber\\
&&\hspace{5cm}\times
C_{12}(0,q^2,M_H^2,M_{S_i}^2,M_{S_i}^2,M_{V_j}^2) 
\Bigg\},
 \n \\
 \n \\
F_{k,R}^{\text{$V_{k}^{0*}$-poles}}|_{S_i,V_j}
&=&F_{k,L}^{\text{$V_{k}^{0*}$-poles}}|_{S_i,V_j}
\Big( g_{V^0_k ff}^{L} \rightarrow
g_{V^0_k ff}^{R}\Big). 
\end{eqnarray}
We also consider the contributions of
one-loop triangle diagrams with exchanging 
a scalar boson $S_{j}$ and two vector 
boson $V_{i}$ in the loop. The Feynman diagrams 
are presented as in Fig.~\ref{SiVVj-dig}.
The corresponding form factors for the above
 diagrams are given:
\begin{eqnarray}
F_{k,L}^{\text{$V_{k}^{0*}$-poles}}|_{V_i,S_j} 
&=&
\dfrac{eQ_V}{8\pi^2}
\sum\limits_{V_i,S_j, V^0_{k}}
\dfrac{g_{HV_iS_j} \; g_{V^0_k V_iS_j}
\,g_{V^0_k ff}^L }{M_H^2 M_{V_i}^2 q^2 (M_H^2-q^2)
(q^2 - M_{V^0_k}^2 + i\Gamma_{V^0_k}M_{V^0_k} )}
\times
\\
&&\hspace{-2cm} \times
\Bigg\{
(M_H^2-q^2) (M_H^2-M_{S_j}^2+M_{V_i}^2) 
\Big[A_0(M_{S_j}^2)-A_0(M_{V_i}^2)\Big]
\n \\
&&\hspace{-2cm}
+
q^2 \Big[ M_H^2(M_H^2 +4 M_{V_i}^2)
-(M_{S_j}^2-M_{V_i}^2)^2
\Big] 
B_0(M_H^2,M_{V_i}^2,M_{S_j}^2)
\n \\
&&\hspace{-2cm}
- M_H^2 \Big[
M_H^2 (M_{S_j}^2-M_{V_i}^2+q^2)
+M_{S_j}^2 (2 M_{V_i}^2-M_{S_j}^2-q^2)
\n \\
&&\hspace{4cm}
+ M_{V_i}^2(5 q^2-M_{V_i}^2)
\Big]
B_0(q^2,M_{V_i}^2,M_{S_j}^2)
\n \\
&&\hspace{-2.0cm}
+2 M_H^2 q^2 (M_H^2-q^2) (M_H^2-M_{S_j}^2+M_{V_i}^2) 
C_{12}(0,q^2,M_H^2,M_{V_i}^2,M_{V_i}^2,M_{S_j}^2)
\n \\
&&\hspace{-2cm}
+4 M_H^2 M_{V_i}^2 q^2 (M_H^2-q^2) 
C_0(0,q^2,M_H^2,M_{V_i}^2,M_{V_i}^2,M_{S_j}^2)
\Bigg\},
 \n \\
 \n \\
F_{k,R}^{\text{$V_{k}^{0*}$-poles}}|_{V_i,S_j}
&=&F_{k,L}^{\text{$V_{k}^{0*}$-poles}}|_{V_i,S_j}
\Big( g_{V^0_k ff}^{L} \rightarrow
g_{V^0_k ff}^{R}\Big). 
\end{eqnarray}
Finally, we have consider fermions exchanging 
in the one-loop triangle diagrams
(shown in Fig.~\ref{fifj-dig}). 
The form factors then read 
\begin{eqnarray}
\label{fifj-poles}
F_{k,L}^{\text{$V_{k}^{0*}$-poles}}|_{f_i,f_j} 
&=&
\dfrac{e Q_f}{4\pi^2}
\sum\limits_{f_i,f_j, V^0_{k}}
\dfrac{N_C^f\; g_{V^0_k ff}^L}
{(q^2 - M_{V^0_k}^2 + i\Gamma_{V^0_k}M_{V^0_k}) }
\times
\\
&&\hspace{-2cm}
\times
\Bigg\{
\Big[
2 m_{f_i} (g_{H f_i f_j}^L g_{V^0_k f_i f_j}^L 
+g_{H f_i f_j}^R g_{V^0_k f_i f_j}^R)
+ 2 m_{f_j} (g_{H f_i f_j}^L g_{V^0_k f_i f_j}^R
+g_{H f_i f_j}^R g_{V^0_k f_i f_j}^L)
\Big] 
\times
\n \\
&&\hspace{-2cm}
\times
\Big[C_{22}+C_{12} \Big]
(0,q^2,M_H^2,m_{f_i}^2,m_{f_i}^2,m_{f_j}^2)
\n \\
&&\hspace{-2.0cm}
+\Big[
3 m_{f_i} (g_{H f_i f_j}^L g_{V^0_k f_i f_j}^L 
+g_{H f_i f_j}^R g_{V^0_k f_i f_j}^R)
+ m_{f_j} (g_{H f_i f_j}^L g_{V^0_k f_i f_j}^R
+g_{H f_i f_j}^R g_{V^0_k f_i f_j}^L)
\Big]  
\times
\n \\
&&\hspace{-2cm}
\times
C_2(0,q^2,M_H^2,m_{f_i}^2,m_{f_i}^2,m_{f_j}^2)
 \n \\
&&\hspace{-2.0cm}
+m_{f_i} (g_{H f_i f_j}^L g_{V^0_k f_i f_j}^R
+g_{H f_i f_j}^R g_{V^0_k f_i f_j}^R) 
C_0(0,q^2,M_H^2,m_{f_i}^2,m_{f_i}^2,m_{f_j}^2)
\Bigg\},
\n \\ 
F_{k,R}^{\text{$V_{k}^{0*}$-poles}}|_{f_i,f_j} 
&=&F_{k,L}^{\text{$V_{k}^{0*}$-poles}}|_{f_i,f_j} 
\Big( g_{V^0_k ff}^{L} \rightarrow
g_{V^0_k ff}^{R}\Big).
\end{eqnarray}
\subsection{Non-$V_{k}^{0*}$ pole contributions} 
We turn our attention to the non-$V_{k}^{0*}$ pole 
contributions, considering all possible 
particles exchanging in the loop diagrams $(c+d)$. 
One first arrives at the group of 
Feynman diagrams with vector boson $V_{i,j}$ internal
lines (as depicted in Fig.~\ref{ViVj-non-dig}).
Analytic formulas for the form factors are given:
\begin{figure}[]
\centering
\includegraphics[width=12cm, height=7cm]
{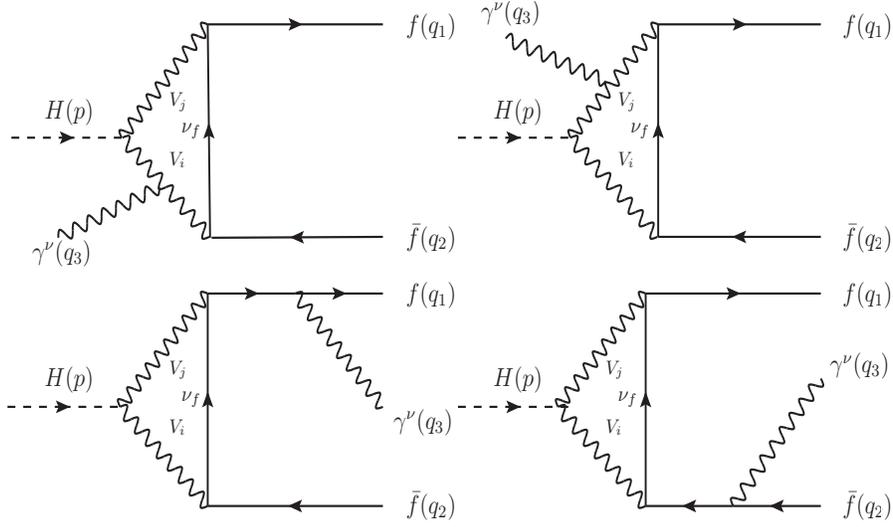}
\caption{\label{ViVj-non-dig} One-loop triangle and box 
diagrams with exchanging vector bosons $V_{i,j}$ in the 
loop.}
\end{figure}
\begin{figure}[]
\centering 
\includegraphics[width=12cm, height=7cm]
{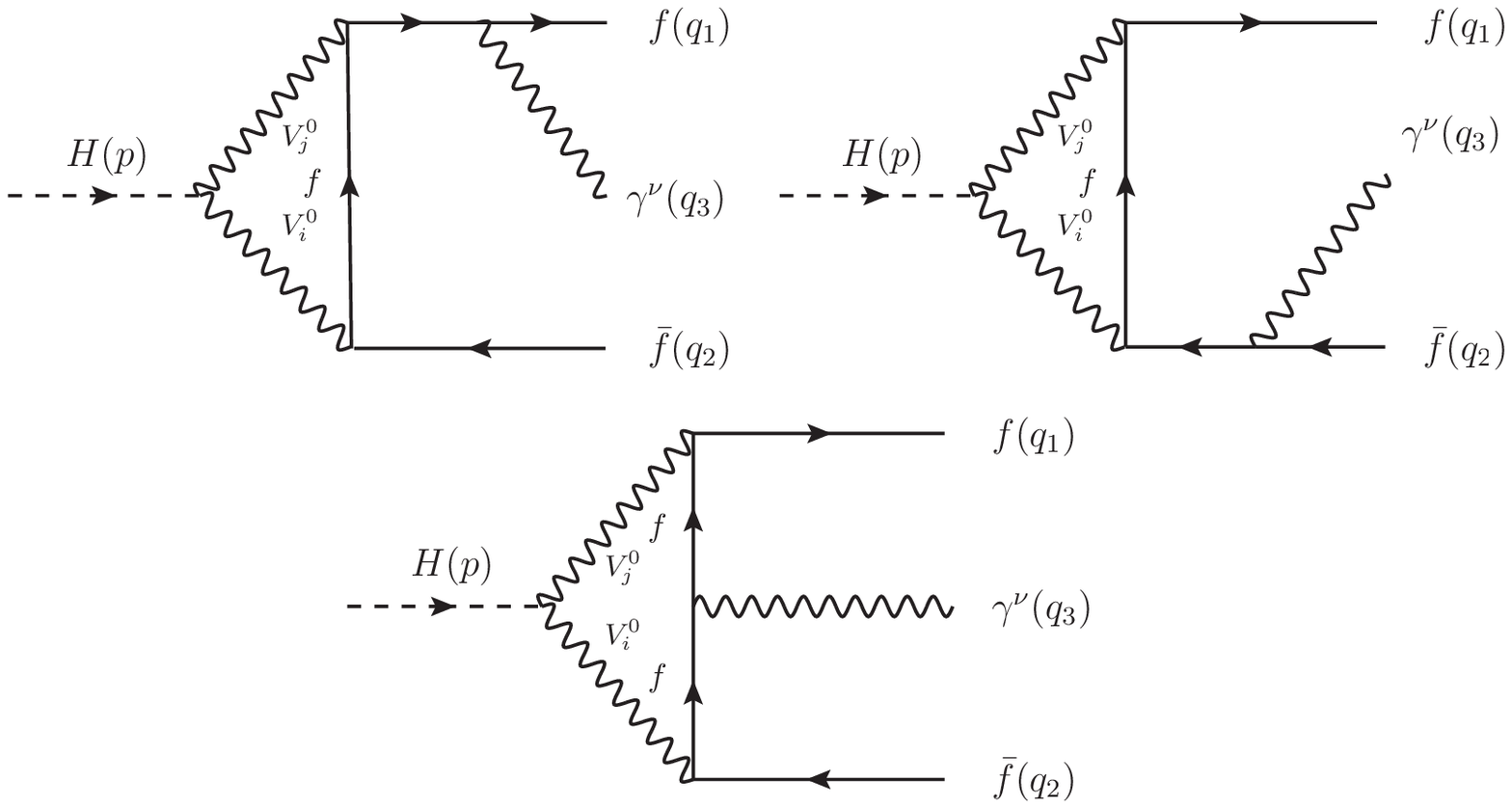}
\caption{\label{V0iV0j-non-dig} One-loop triangle and 
box diagrams with exchanging vector bosons 
$V^0_{i}, V^0_{j}$ in 
the loop.}
\end{figure}
\begin{figure}[]
\centering
\includegraphics[width=12cm, height=7cm]
{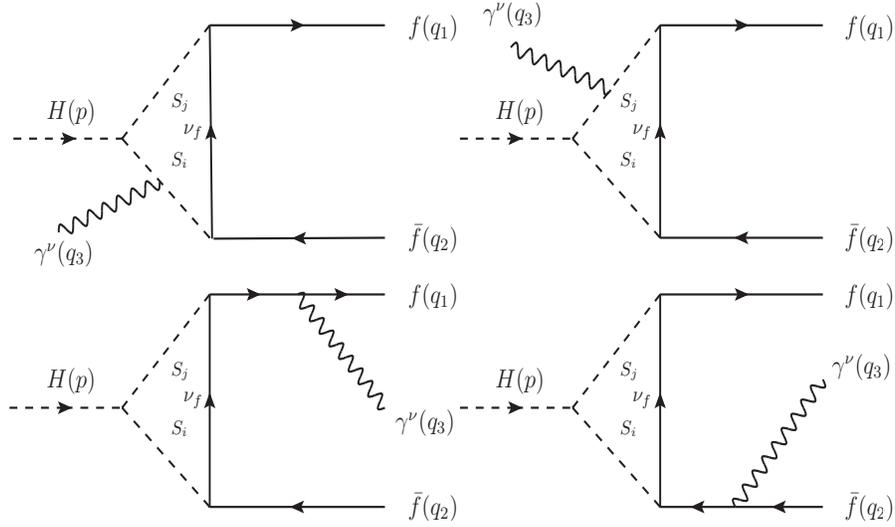}
\caption{\label{SiSj-non-dig} One-loop diagrams with 
exchanging
charged scalar bosons $S_{i,j}$ in the loop.}
\end{figure}
\begin{figure}[]
\centering
\includegraphics[width=12cm, height=7cm]
{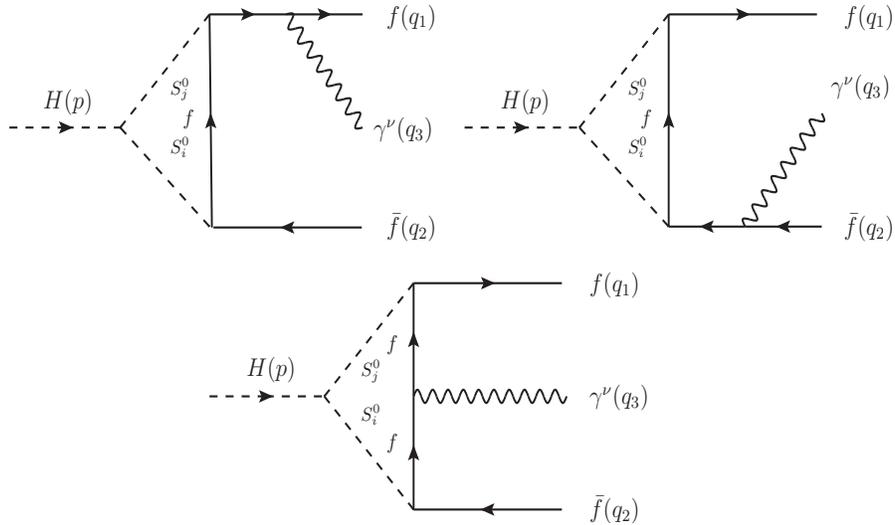}
\caption{\label{S0iS0j-non-dig} One-loop box diagrams 
with exchanging vector boson $S^0_{i,j}$ 
particles in loop.}
\end{figure}
\begin{figure}[]
\centering
\includegraphics[width=12cm, height=7cm]
{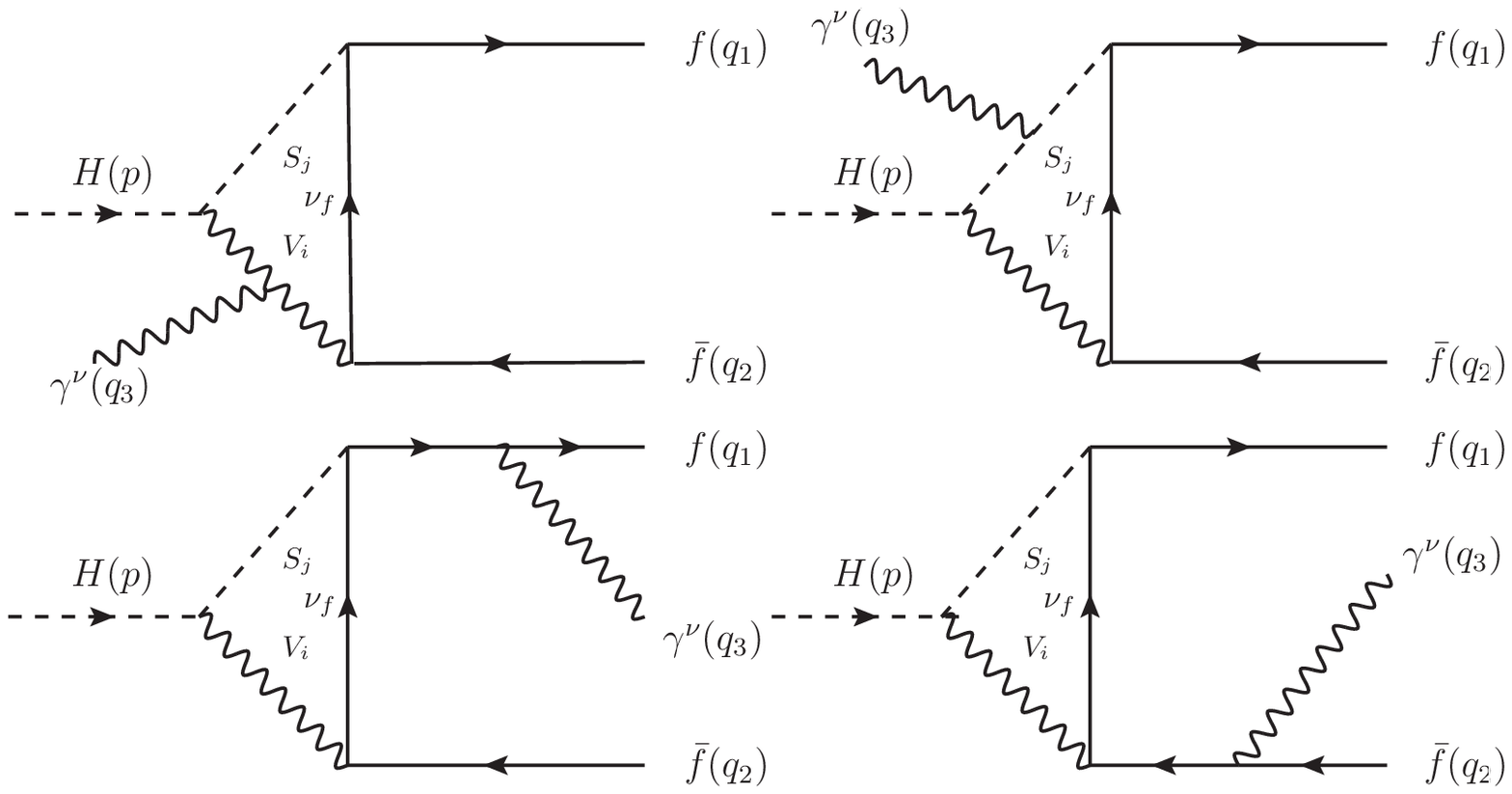}
\\
\includegraphics[width=12cm, height=7cm]
{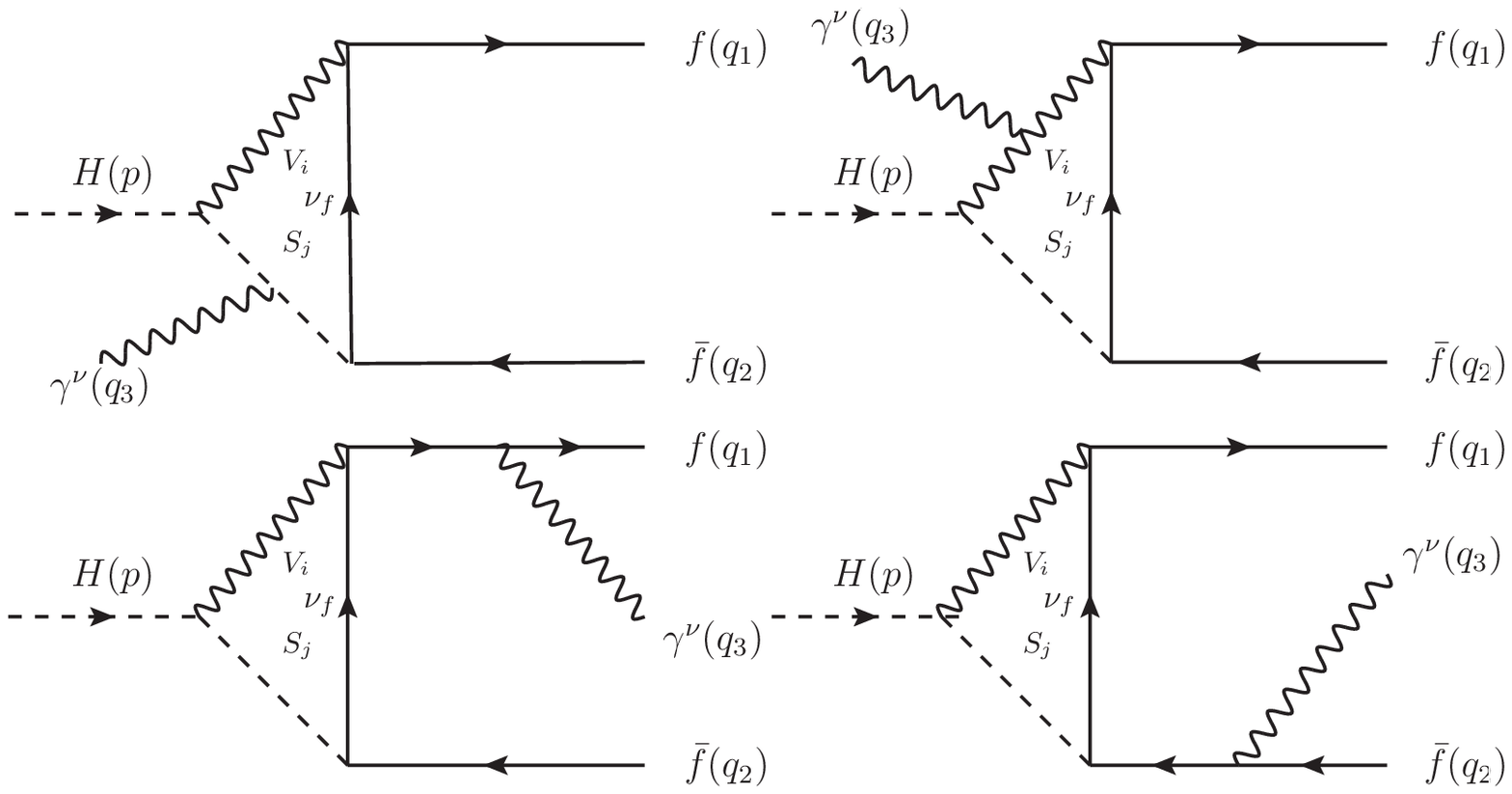}
\caption{\label{ViSj-non-dig} One-loop diagrams 
with exchanging
vector boson $V_{i}$ and scalar boson $S_j$
particles in loop.}
\end{figure}
\begin{figure}[]
\centering
\includegraphics[width=12cm, height=7cm]
{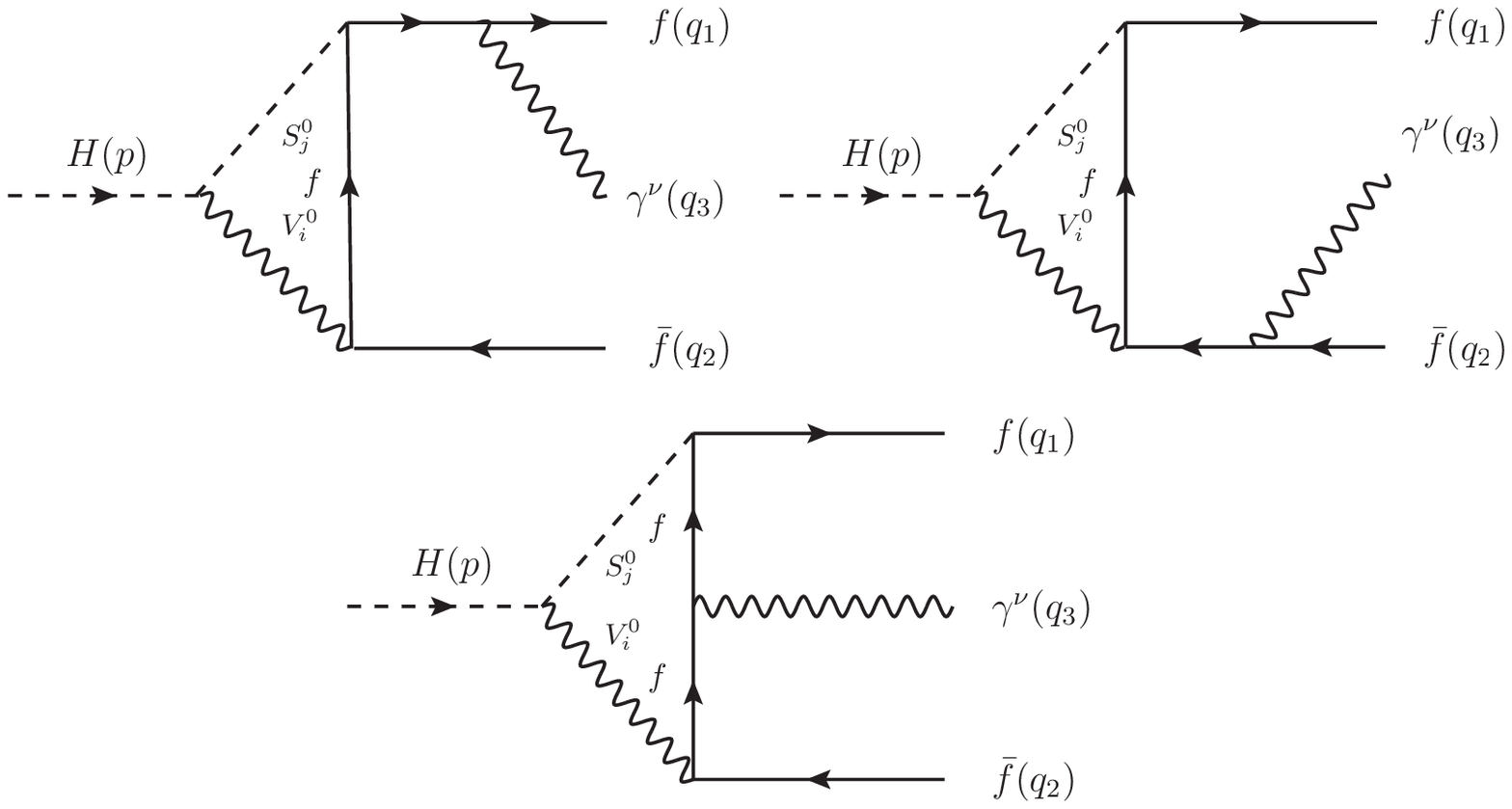}
\\
\includegraphics[width=12cm, height=7cm]
{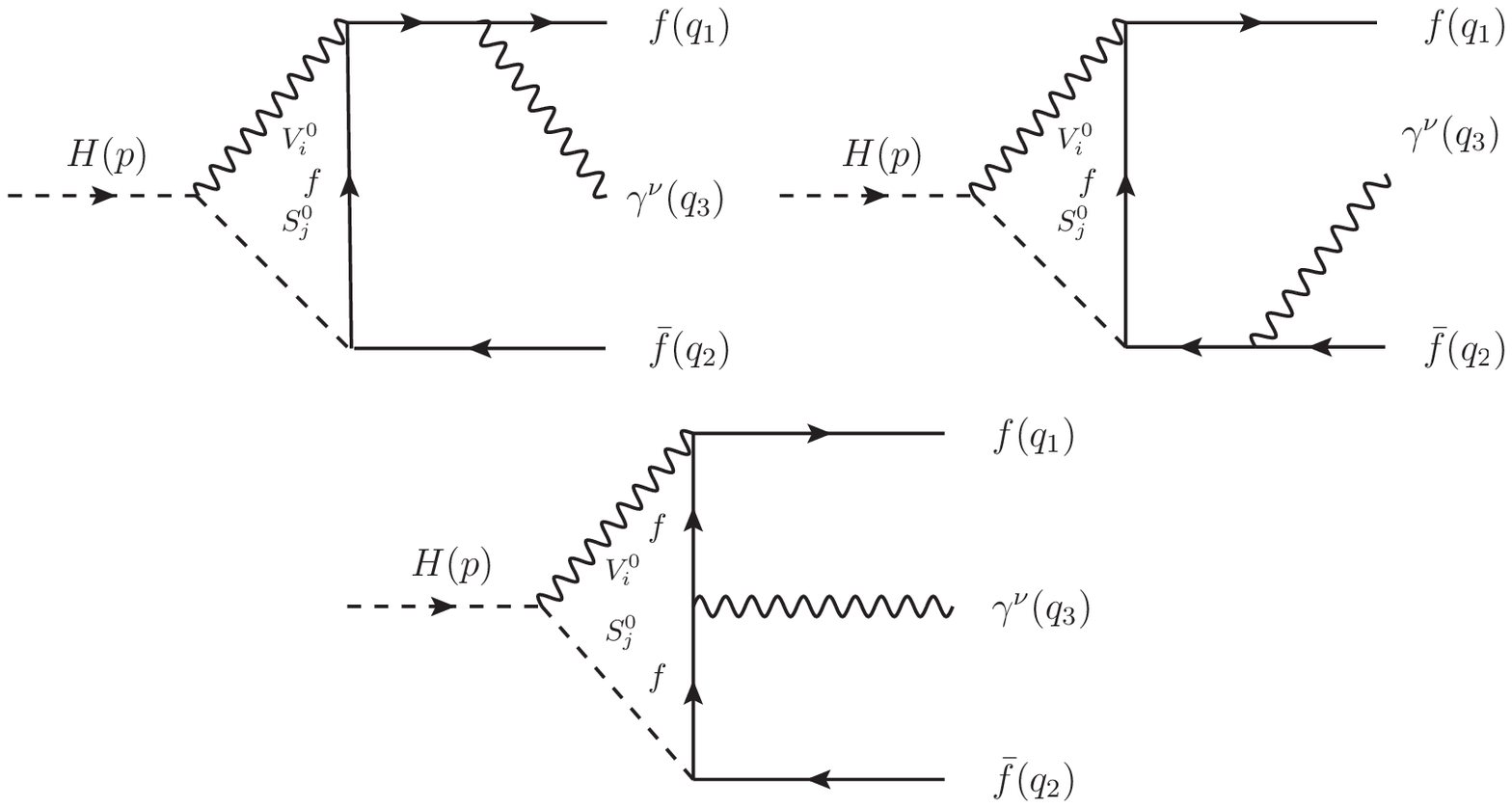}
\caption{\label{V0iS0j-non-dig} One-loop diagrams 
with exchanging
vector boson $V^0_{i}$ and scalar boson $S_j^0$ 
particles in loop.}
\end{figure}
\begin{eqnarray}
\label{ViVj-nonPoles}
F_{1,L}^{\text{Non-$V_{k}^{0*}$} }|_{V_i,V_j} 
&=& \dfrac{eQ_V}{16\pi^2}
\sum\limits_{V_i,V_j}
\dfrac{ g_{HV_iV_j} \, g_{V_i f \nu_f}^L
\, g_{V_j f \nu_f}^L}{M_{V_i}^2 M_{V_j}^2}
\times
\n \\
&&\hspace{-0.5cm} \times
\Bigg\{
- 2 M_{V_i}^2 
\Big[
C_0(0,q_{12},M_H^2,M_{V_i}^2,M_{V_i}^2,M_{V_j}^2)
+
C_0(q_{12},0,M_H^2,M_{V_i}^2,M_{V_j}^2,M_{V_j}^2)
\Big]
\n \\
&&\hspace{-.5cm}
-
(M_H^2+M_{V_i}^2+M_{V_j}^2) 
\Big\{
\Big[C_{22}+C_{12}\Big]
(0,q_{12},M_H^2,M_{V_i}^2,M_{V_i}^2,M_{V_j}^2)
\n \\
&&\hspace{5cm}
+
\Big[C_{22}+C_{12}\Big]
(q_{12},0,M_H^2,M_{V_i}^2,M_{V_j}^2,M_{V_j}^2)
\Big\}
\n \\
&&\hspace{-.5cm}
- (M_H^2+3 M_{V_i}^2 - M_{V_j}^2) 
\Big[
C_2(0,q_{12},M_H^2,M_{V_i}^2,M_{V_i}^2,M_{V_j}^2)
\n \\
&&\hspace{5.5cm}
+C_2(q_{12},0,M_H^2,M_{V_i}^2,M_{V_j}^2,M_{V_j}^2)
\Big]
\n \\
&&\hspace{-.5cm}
+2  (M_{V_j}^2 - M_{V_i}^2) 
C_1(q_{12},0,M_H^2,M_{V_i}^2,M_{V_j}^2,M_{V_j}^2)
\\
&&\hspace{-.5cm}
+(8 - 2 d) M_{V_i}^2 M_{V_j}^2 
\Big[
D_3(0,0,0,M_H^2;q_{12},q_{13},M_{V_i}^2,0,M_{V_j}^2,M_{V_j}^2)
\n \\
&&\hspace{3.5cm}
+D_3(0,0,0,M_H^2;q_{23},q_{12},M_{V_i}^2,M_{V_i}^2,0,M_{V_j}^2))
\Big]
\n \\
&&\hspace{-.5cm}
+(4 - 2 d) M_{V_i}^2 M_{V_j}^2
\Big\{
\Big[D_{33} + D_{23} + D_{13}\Big]
(0,0,0,M_H^2;q_{23},q_{12},M_{V_i}^2,M_{V_i}^2,0,M_{V_j}^2)
\n \\
&&\hspace{3.5cm}
+ \Big[D_{33} + D_{23} \Big]
(0,0,0,M_H^2;q_{12},q_{13},M_{V_i}^2,0,M_{V_j}^2,M_{V_j}^2)
\Big\}
\n \\
&&\hspace{-.5cm}
+ 4 M_{V_i}^2 M_{V_j}^2
D_2(0,0,0,M_H^2;q_{12},q_{13},M_{V_i}^2,0,M_{V_j}^2,M_{V_j}^2)
\Bigg\},
\n \\
\n \\
F_{1,R}^{\text{Non-$V_{k}^{0*}$}}|_{V_i,V_j}
&=& F_{1,L}^{\text{Non-$V_{k}^{0*}$}}|_{V_i,V_j} 
\Big(g_{V_i f \nu_f}^L \rightarrow
g_{V_i f \nu_f}^R ; 
 \; g_{V_j f \nu_f}^{L}  \rightarrow
g_{V_j f \nu_f}^{R} 
\Big),\\
F_{2,L}^{\text{Non-$V_{k}^{0*}$}}|_{V_i,V_j}
&=& 
F_{1,L}^{\text{Non-$V_{k}^{0*}$}}|_{V_i,V_j} 
\Big(
\{q_{13}, q_{23}\}
\rightarrow 
\{q_{23}, q_{13}\}
\Big),
\\
F_{2,R}^{\text{Non-$V_{k}^{0*}$}}|_{V_i,V_j} 
&=& F_{2,L}^{\text{Non-$V_{k}^{0*}$}}|_{V_i,V_j}
\Big(g_{V_i f \nu_f}^L \rightarrow
g_{V_i f \nu_f}^R; 
 \; g_{V_j f \nu_f}^{L}  \rightarrow
g_{V_j f \nu_f}^{R} 
\Big).
\end{eqnarray}
We find that the result is presented in terms of 
$C$- and $D$-functions up to $D_{33}$-coefficients.
The reason for this fact is explained as follows. 
Due to the exchange of vector boson in the loop, 
we have to handle with tensor one-loop integrals with 
rank $R\geq 3$ in the amplitude 
of each diagram. However, they are cancelled out 
after summing all diagrams. As a result, 
the total amplitude is only expressed in terms 
of tensor integrals with $R \leq 2$ that causes 
of leading the results.

For neutral vector boson $V^0_{i}, V^0_{j}$ 
internal lines in the loop diagrams 
(seen Fig.~\ref{V0iV0j-non-dig}), the 
corresponding form factors are obtained 
\begin{eqnarray}
\label{V0iV0j-nonPloes}
F_{1,L}^{\text{Non-$V_{k}^{0*}$} }|_{V^0_i,V^0_j} 
&=&
\dfrac{e Q_f}{16 \pi ^2}
\sum\limits_{V^0_i,V^0_j}
g_{HV^0_i V^0_j}\; 
g_{V^0_i f f}^L \;
g_{V^0_j f f}^L
\times
\\
&&\hspace{-0.5cm} \times
\Big\{
(4-2 d)
\big[D_{33}+D_{23}\big]
+
(8-2 d) D_3
\Big\}
(0,0,0,M_H^2,q_{23},q_{13},M_{V^0_i}^2,0,0,M_{V^0_j}^2),
\n \\
\n \\
F_{1,R}^{\text{Non-$V_{k}^{0*}$}}|_{V^0_i,V^0_j}
&=& F_{1,L}^{\text{Non-$V_{k}^{0*}$}}|_{V^0_i,V^0_j} 
\Big(g_{V^0_i f f}^L \rightarrow
g_{V^0_i f f}^R ; 
 \; g_{V^0_j f f}^{L}  \rightarrow
g_{V^0_j f f}^{R} 
\Big),\\
F_{2,L}^{\text{Non-$V_{k}^{0*}$}}|_{V^0_i,V^0_j}
&=& 
F_{1,L}^{\text{Non-$V_{k}^{0*}$}}|_{V^0_i,V^0_j} 
\Big(
\{q_{13}, q_{23}\}
\rightarrow 
\{q_{23}, q_{13}\}
\Big),
\\
F_{2,R}^{\text{Non-$V_{k}^{0*}$}}|_{V^0_i,V^0_j} 
&=& F_{2,L}^{\text{Non-$V_{k}^{0*}$}}|_{V^0_i,V^0_j}
\Big(g_{V^0_i f f}^L \rightarrow
g_{V^0_i f f}^R ; 
 \; g_{V^0_j f f}^{L}  \rightarrow
g_{V^0_j f f}^{R} 
\Big).
\end{eqnarray}
Next, we also consider one-loop diagrams 
with having charged scalar bosons $S_{i,j}$
internal lines (shown in Fig.~\ref{SiSj-non-dig}). 
The results read
\begin{eqnarray}
F_{1,L}^{\text{Non-$V_{k}^{0*}$} }|_{S_i,S_j} 
&=&
\dfrac{eQ_S}{8 \pi ^2}
\sum\limits_{S_i,S_j}
g_{HS_iS_j} \, g_{S_i f \nu_f}^R \, g_{S_j f \nu_f}^R
\times
\n \\
&&\hspace{0.0cm} \times
\Bigg\{
\Big[D_{33}+D_{23}+D_3 \Big]
(0,0,0,M_H^2;q_{12},q_{13},M_{S_i}^2,0,M_{S_j}^2,M_{S_j}^2)
\\
&&\hspace{.3cm}
+\Big[D_{33}+D_{23}+D_{13}+D_3 \Big]
(0,0,0,M_H^2;q_{23},q_{12},M_{S_i}^2,
M_{S_i}^2,0,M_{S_j}^2)
\Bigg\}
\n ,
\\
F_{1,R}^{\text{Non-$V_{k}^{0*}$}}|_{S_i,S_j}
&=& F_{1,L}^{\text{Non-$V_{k}^{0*}$}}|_{S_i,S_j} 
\Big(g_{S_i f \nu_f}^R \rightarrow
g_{S_i f \nu_f}^L ; 
 \; g_{S_j f \nu_f}^R  \rightarrow
g_{S_j f \nu_f}^L
\Big),\\
F_{2,L}^{\text{Non-$V_{k}^{0*}$}}|_{S_i,S_j}
&=& 
F_{1,L}^{\text{Non-$V_{k}^{0*}$}}|_{S_i,S_j} 
\Big(
\{q_{13}, q_{23}\}
\rightarrow 
\{q_{23}, q_{13}\}
\Big),
\\
F_{2,R}^{\text{Non-$V_{k}^{0*}$}}|_{S_i,S_j} 
&=& F_{2,L}^{\text{Non-$V_{k}^{0*}$}}|_{S_i,S_j}
\Big(g_{S_i f \nu_f}^R \rightarrow
g_{S_i f \nu_f}^L ; 
 \; g_{S_j f \nu_f}^R  \rightarrow
g_{S_j f \nu_f}^L
\Big).
\end{eqnarray}

Furthermore, 
one also has the contributions of 
neutral scalar boson $S^0_{i,j}$ 
exchanging in the loop diagrams (as described in
Fig.~\ref{S0iS0j-non-dig}).
Analytic results for the form factors then 
read
\begin{eqnarray}
F_{1,L}^{\text{Non-$V_{k}^{0*}$} }|_{S^0_i,S^0_j} 
&=&
- \dfrac{e Q_f}{8\pi ^2}\sum\limits_{S^0_i,S^0_j}
\, g_{HS^0_i S^0_j} \, 
g_{S^0_i ff}^L \, g_{S^0_j ff}^R 
\times
\\
&&\hspace{0.0cm} \times
\Big[D_{33}+D_{23}+D_3 \Big]
(0,0,0,M_H^2;q_{23},q_{13};M_{S^0_i}^2,0,0,M_{S^0_j}^2)
\n ,
\\
F_{1,R}^{\text{Non-$V_{k}^{0*}$}}|_{S^0_i,S^0_j}
&=& F_{1,L}^{\text{Non-$V_{k}^{0*}$}}|_{S^0_i,S^0_j} 
\Big(g_{S^0_i ff}^L \rightarrow
g_{S^0_i ff}^R ; 
 \; g_{S^0_j ff}^R \rightarrow
g_{S^0_j ff}^L
\Big),\\
F_{2,L}^{\text{Non-$V_{k}^{0*}$}}|_{S^0_i,S^0_j}
&=& 
F_{1,L}^{\text{Non-$V_{k}^{0*}$}}|_{S^0_i,S^0_j} 
\Big(
\{q_{13}, q_{23}\}
\rightarrow 
\{q_{23}, q_{13}\}
\Big),
\\
F_{2,R}^{\text{Non-$V_{k}^{0*}$}}|_{S^0_i,S^0_j} 
&=& F_{2,L}^{\text{Non-$V_{k}^{0*}$}}|_{S^0_i,S^0_j}
\Big(g_{S^0_i ff}^L \rightarrow
g_{S^0_i ff}^R ; 
 \; g_{S^0_j ff}^R \rightarrow
g_{S^0_j ff}^L
\Big).
\end{eqnarray}
We now consider non-$V_k^{*0}$ pole 
one-loop diagrams with mixing of scalar
$S_j$ (or $S_j^0$) and $V_i (V^0_i)$
exchanging in the loop. The diagrams are 
depicted in Figs.~\ref{ViSj-non-dig},\;
\ref{V0iS0j-non-dig}. 
The calculations are performed as same procedure.
We finally find that these contributions 
are proportional to $m_f$. As a result, 
in the limit of $m_f \rightarrow 0$, one confirms 
that
\begin{eqnarray}
F_{k,L}^{\text{Non-$V_{k}^{0*}$} }|_{V_i,S_j} 
=
F_{k,R}^{\text{Non-$V_{k}^{0*}$} }|_{V_i,S_j} 
&=& 0, \\
F_{k,L}^{\text{Non-$V_{k}^{0*}$} }|_{V^0_i,S^0_j} 
=
F_{k,R}^{\text{Non-$V_{k}^{0*}$} }|_{V^0_i,S^0_j} 
&=& 0. 
\end{eqnarray}
for $k = 1,2$.

We verify the ultraviolet finiteness of the results. 
We find that the UV-divergent parts 
of all the above form factors come from all $B$-functions. 
While $C$- and $D$-functions in this papers are UV-finite. 
Higher rank tensor $B$-functions can be reduced into 
$B_0$ and $A_0$. We verify that the sum of all 
$B$-functions gives a UV-finite result. As a result, 
all the form factors are UV-finite (seen our previous 
paper~\cite{Phan:2021xwc} for example).

Having the correctness form 
factors for the decay processes, the decay rate 
is given by~\cite{Kachanovich:2020xyg}: 
\begin{eqnarray}
\label{decayrate}
 \dfrac{d\Gamma}{d q_{12}q_{13}} 
 = \dfrac{q_{12}}{512 \pi^3 M_H^3}
 \Big[
 q_{13}^2(|F_{1,R}|^2 + |F_{2,R}|^2)
 + q_{23}^2(|F_{1,L}|^2 + |F_{2,L}|^2)
 \Big]. 
\end{eqnarray}
Taking the above integrand over $0 \leq q_{12} \leq M_H^2$ 
and $0\leq q_{13} \leq M_H^2 - q_{12}$, one gets the 
total decay rates. In the next subsection, we show 
typical examples which we apply our analytical results 
for $H\rightarrow f \bar{f} \gamma$ to the Standard 
Model, the $U(1)_{B-L}$ extension of the SM, THDM. 
Phenomenological results for the
decay channels of the mentioned models
also studied with using 
the present parameters at the LHC. 
\section{Applications}  
We are going to apply the above 
results to the standard model
and several BSMs such as  the 
$U(1)_{B-L}$ extension of the SM, 
THDM. For phenomenological analyses, 
we use the following input parameters:
$\alpha=1/137.035999084$, 
$M_Z = 91.1876$ GeV, 
$\Gamma_Z  = 2.4952$ GeV, 
$M_W = 80.379$ GeV, $M_H =125.1$ GeV,
$m_{\tau} = 1.77686$ GeV, $m_t= 172.76$ GeV, 
$m_b= 4.18$ GeV, $m_s = 0.93$ GeV 
and $m_c = 1.27$ GeV. 
Depending on the models under consideration,
the input values for new parameters are then 
shown. 
\subsection{Standard model}
We first reduce our result to the case of
the standard model. In this case, we have $V_i, 
V_j \rightarrow W^+, W^-$, 
$V_k^0 \rightarrow Z, \gamma$. 
All couplings relating to the decay channels 
$H\rightarrow f\bar{f} \gamma$ in the SM
are replaced as in Table~\ref{tableSM}.
\begin{table}
\centering
{\begin{tabular}{l@{\hspace{5cm}}l }
\hline \hline
\textbf{Vertices} & \textbf{Couplings}\\
\hline \hline 
$g_{HV_iV_j}$            &   $eM_W/s_W$               \\ \\
$g_{V^0_kV_iV_j}$        &   $e \; c_W/s_W$           \\ \\
$g_{V^0_kAV_iV_j}$       &   $e^2 \; c_W/s_W$         \\ \\
$g_{V^0_k \nu_f\nu_f}^L$ &   $e/(2 s_W c_W) $         \\ \\
$g_{V^0_k\nu_f\nu_f}^R $ &   $0$                      \\ \\
$ g_{Hf_if_j}^{L, R}$    &   $e \; m_f/(2 s_W\; M_W)$ \\ \\
$g_{V^0_kf_if_j}^L$      &   $e(T_{3}^f - Q_f\; s^2_W)/(s_W \; c_W)$ 
\\ \\
$g_{V^0_kf_if_j}^R$      &   $-e Q_f s_W/c_W$ \\ \\
$g_{V_i f\nu_f}^L$       &   $e/(\sqrt{2}\;s_W)$ \\ \\
$g_{V_i f\nu_f}^R$       &   $0$\\ \hline \hline
\end{tabular}}
\caption{
\label{tableSM}
All the couplings involving the decay processes
$H\rightarrow f\bar{f} \gamma$ in the SM. 
}
\end{table}
In the SM, the form factors are obtained by taking 
the contributions of~Eqs.~(\ref{ViVj-poles},
\ref{fifj-poles}, \ref{ViVj-nonPoles},
\ref{V0iV0j-nonPloes}). Using the above couplings, 
we then get a compact expression for the form factors 
as follows:
\begin{eqnarray}
F_{2L} &=&
\dfrac{\alpha^2 m_t^2}{3 s_W M_W}
\Bigg[ \dfrac{16}{q_{12}}
+
\dfrac{2(8s_W^2-3)}{e s_W c_W}\;
\dfrac{g_{Zff}^L}{q_{12}- M_Z^2+ i \Gamma_Z M_Z}
\Bigg] \times
\nonumber \\
&&\hspace{0.7cm} 
\times
\Bigg\{
4 \Big(C_{22}+ C_{12}+ C_2\Big) + C_0 \Bigg\}
(0,q_{12},M_H^2,m_t^2,m_t^2,m_t^2)
\nonumber\\
&& +
\dfrac{\alpha^2}{ s_W M_W^3}
\Bigg[ 
\dfrac{1}{q_{12}}
- \dfrac{c_W}{e s_W}
\dfrac{g_{Zff}^L}{q_{12}- M_Z^2+ i \Gamma_Z M_Z}
\Bigg]
\times
\nonumber\\
&&\hspace{0.7cm} 
\times 
\Bigg\{
\Big[ 
-M_H^2 
\Big( B_1 + 3 B_{11} + 2 B_{111} \Big)
- 2 \Big(
B_{00} + 2 B_{001} 
\Big) \Big] (M_H^2,M_W^2,M_W^2)
\nonumber \\
&&\hspace{0cm}
+\Big[
4 M_W^2 
\Big( q_{12}-M_H^2 \Big)
+ 2 M_H^2 q_{12} 
+ 8 (1-d) M_W^4
\Big]  \times
\nonumber \\
&&
\hspace{3.0cm} 
\times
\Big(
C_{22}+ C_{12}+ C_2
\Big)
(0,q_{12},M_H^2,M_W^2,M_W^2,M_W^2)
\nonumber \\
&&
\hspace{0cm}
+ 4 M_W^2 (q_{12}-4 M_W^2) 
C_0(0,q_{12},M_H^2,M_W^2,M_W^2,M_W^2)
\Bigg\}
\nonumber\\
&&
+ \dfrac{\alpha}{4\pi s_W M_W^3}
\Big( g_{Wf \nu_f}^L  \Big)^2
\times 
\\
&& \hspace{0.7cm} 
\times
\Bigg\{
(-M_H^2-2 M_W^2) 
\Big[
\Big( 
C_{12} + C_{11} + C_1 
\Big)
(M_H^2,0,q_{12},M_W^2,M_W^2,M_W^2)
\nonumber \\
&&
\hspace{4.3cm} 
+\Big( 
C_{12} + C_{11} + C_1 
\Big)
(M_H^2,q_{12},0,M_W^2,M_W^2,M_W^2)
\Big]
\nonumber \\
&&
\hspace{0cm} 
-2 M_W^2 
\Big[
C_0(M_H^2,0,q_{12},M_W^2,M_W^2,M_W^2)
+ C_0(M_H^2,q_{12},0,M_W^2,M_W^2,M_W^2)
\Big]
\nonumber \\
&&
\hspace{0cm} 
+ 2M_W^4 (2 - d ) 
\Big[
\Big(
D_{13}+D_{12}+D_{11} 
\Big) 
(M_H^2,0,0,0;q_{12},q_{13},M_W^2,M_W^2,M_W^2,0)
\nonumber \\
&&
\hspace{0.0cm} 
+\Big( 
D_{23}+D_{22}+D_{13}+2 D_{12}+D_{11} 
\Big)
(M_H^2,0,0,0;q_{23},q_{12},M_W^2,M_W^2,0,M_W^2)
\Big]
\nonumber \\
&&
\hspace{0cm} 
-4 M_W^4 
\Big[
\Big(
D_3+ D_2+ D_0
\Big)
(M_H^2,0,0,0;q_{12},q_{13},M_W^2,M_W^2,M_W^2,0)
\nonumber \\
&&\hspace{4.0cm} 
+ D_0(M_H^2,0,0,0;q_{23},q_{12},M_W^2,M_W^2,0,M_W^2)
\Big]
\nonumber \\
&&\hspace{0cm} 
-2 d M_W^4 
\Big[
\Big(
D_2+ D_1 
\Big)
(M_H^2,0,0,0;q_{23},q_{12},M_W^2,M_W^2,0,M_W^2)
\nonumber \\
&&\hspace{4.0cm} 
+D_1(M_H^2,0,0,0;q_{12},q_{13},M_W^2,M_W^2,M_W^2,0)
\Big]
\Bigg\}
\nonumber\\
&&
+
\dfrac{\alpha M_W}{2\pi c_W^2 s_W}
\Big(g_{Zf f}^L\Big)^2 
\Bigg\{
(d-2)\times
\nonumber\\
&&
\hspace{1cm}
\times 
\Big[
D_{23}+D_{22}+D_{13}+2D_{12}+D_{11} 
\Big]
(M_H^2,0,0,0;q_{23},q_{13},M_Z^2,M_Z^2,0,0)
\nonumber \\
&&\hspace{1.cm} 
+\Big[ d 
\Big(
D_2+D_1
\Big)
+2 
\Big(
D_3+D_0
\Big)
\Big] 
(M_H^2,0,0,0;q_{23},q_{13},M_Z^2,M_Z^2,0,0)
\Bigg\}. \nonumber
\end{eqnarray}
Where some coupling constants relate in this 
representation like
$g_{Wf \nu_f}^L = e/(\sqrt{2} s_W)$, 
$g_{Zf f}^L = e (2 s_W^2 -1)/(2 c_W s_W)$ and 
$g_{Wf \nu_f}^R = 0$, $g_{Zf f}^R = e s_W /c_W$. 

Other form factors can be obtained 
as follows. 
\begin{eqnarray}
F_{1L}
&=& 
F_{2L}
\Big(
\{q_{13}, q_{23}\}
\rightarrow 
\{q_{23}, q_{13}\}
\Big), \\
F_{kR}
&=& F_{kL}
\Big(
g_{Wf \nu_f}^L
\rightarrow
g_{Wf \nu_f}^R; 
g_{Z f f}^L 
\rightarrow
g_{Zf f}^R 
\Big)
\end{eqnarray}
for $k=1,2$.

It is stress that we derive alternative results 
for the form factors of $H\rightarrow f\bar{f}\gamma$ 
in the SM in comparison with previous works. 
Because our analytic results in this paper are 
computed in the unitary. Thus, our formulas may get 
different forms in comparison with the results 
in~\cite{Kachanovich:2020xyg} which have 
calculated in $R_{\xi}$-gauge. In this paper, 
we perform cross-check our results with 
\cite{Kachanovich:2020xyg} by numerical tests. 
The numerical results for this check
are shown in Table~\ref{NumF1}. We find that 
our results are in good agreement with 
the ones in \cite{Kachanovich:2020xyg} 
with more than $10$ digits. 
\begin{table}[h]
\begin{center}
\begin{tabular}{l@{\hspace{1.5cm}}l@{\hspace{1.5cm}}l}  
\hline \hline 
$( q_{12} , q_{13})$
& This work & Ref.~\cite{Kachanovich:2020xyg}  
\\  \hline \hline\\
$( 100 , 200 )$ & $9.62231539663501 \cdot 10^{-8}$ & $9.62231539662956  \cdot 10^{-8}$ \\ 
& $-3.501515874673991 \cdot 10^{-10} \, i$ & $-3.501515874674078 \cdot 10^{-10} \, i$  
  \\ \hline\\
$( -100 , 200 )$ & $-9.95227151085161 \cdot 10^{-8}$ & $-9.95227151084899 \cdot 10^{-8}$  \\ 
  & $-3.531494528007124 \cdot 10^{-10} \, i$ & $-3.531494528006995 \cdot 10^{-10} \, i$  
  \\ \hline\\
$( 100 , -200 )$ & $9.62360254230002 \cdot 10^{-8}$ & $9.62360254229717 \cdot 10^{-8}$  \\ 
  & $-3.597907189717628 \cdot 10^{-10} \, i$ & $-3.597907189717582 \cdot 10^{-10} \, i$  
  \\ \hline\\
$( -100 , -200 )$ & $-9.95098785515085 \cdot 10^{-8}$ &  $-9.95098785514946\cdot 10^{-8}$  \\ 
  & $-3.622848558573573 \cdot 10^{-10} \, i$ & $-3.622848558573332 \cdot 10^{-10} \, i$      
\\ \hline\hline
\end{tabular}
\caption{\label{NumF1} Numerical checks for form factor $F_{2L}$ 
in this work with $b_1$ in $A2$ of \cite{Kachanovich:2020xyg}. }
\end{center}
\end{table}

We also generate the decay widths for 
$H\rightarrow e\bar{e}\gamma$ and cross-check our 
results with~\cite{Abbasabadi:1995rc}. 
The results are presented in Table~\ref{decaySM}.
For this test, we adjust the input parameters 
and apply all cuts as same 
in~\cite{Abbasabadi:1995rc}.  In the 
Table~\ref{decaySM}, the parameter $k$ is 
taken account which come from the kinematical
cuts of the invariant masses $m_{ff},\; m_{f\gamma}$
as follows:
\begin{eqnarray}
 m_{ff}^2, m_{f\gamma}^2  \geq (k M_H)^2. 
\end{eqnarray}
We find that our results are in good agreement
with the ones in~\cite{Abbasabadi:1995rc}.
\begin{table}
\begin{center}
\begin{tabular}{l@{\hspace{2cm}}l@{\hspace{2cm}}l}  
\hline \hline 
$k$
& This work   & Ref.~\cite{Abbasabadi:1995rc}  
\\ \hline \hline\\
$0$ & $0.576865$ & $0.5782$ \\ \hline \\
$0.1$ & $0.242514$ & $0.245$ \\  \hline \\
$0.2$ & $0.184121$ & $0.1897$ \\  \hline \\
$0.3$ & $0.121368$ & $0.1242$ \\  \hline \\
$0.4$ & $0.0572478$ & $0.05844$ \\  \hline\hline
\end{tabular}
\caption{\label{decaySM} Cross-check the results of 
the decay widths in this work 
with~\cite{Abbasabadi:1995rc}}
\end{center}
\end{table}
\subsection{$U(1)_{B-L}$ extension of the SM}
We refer to the appendix $B$ for reviewing 
briefly this model. In the appendix $B$, all 
couplings relating to the decay processes
$H\rightarrow f\bar{f}\gamma$ are derived in 
$U(1)_{B-L}$ extension of the SM 
(seen Table~\ref{U1-coupling}). 
Apart from all particles in the SM, 
two additional neutral Higgs and 
a neutral gauge boson $Z'$ 
which belongs to $U(1)_{B-L}$ gauge symmetry
are taken into account in this model.

For phenomenological study, we have to include three 
new parameters such as the mixing angle $c_{\alpha}$, 
the $U(1)_{B-L}$ coupling $g'_1$ and the mass of new 
gauge boson $M_Z'$. In all the below results, we 
set $c_{\alpha}=0.3, 0.7$ and $c_{\alpha}=1$
(this case is to the standard model). The mass of $Z'$ 
is in the range of $600$ GeV $\leq M_Z' \leq 1000$ GeV. 
The coupling $g'_1$ is in $0.05 \leq g_1' \leq 0.5$.

\begin{figure}[]
\centering
$ \begin{array}{cc}
\hspace{-8cm}
\dfrac{d\Gamma}{dm_{f\bar{f}}} & 
\hspace{-8cm}
\dfrac{d\Gamma}{dm_{f\gamma} }\\
& \\
\hspace{0cm}
\includegraphics[width=8cm, height=8cm]
{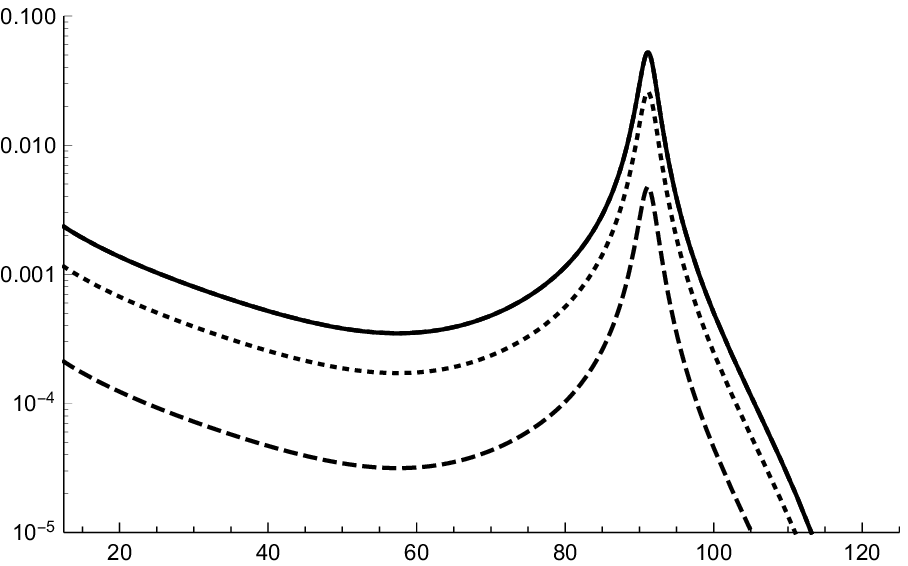}
&
\includegraphics[width=8cm, height=8cm]
{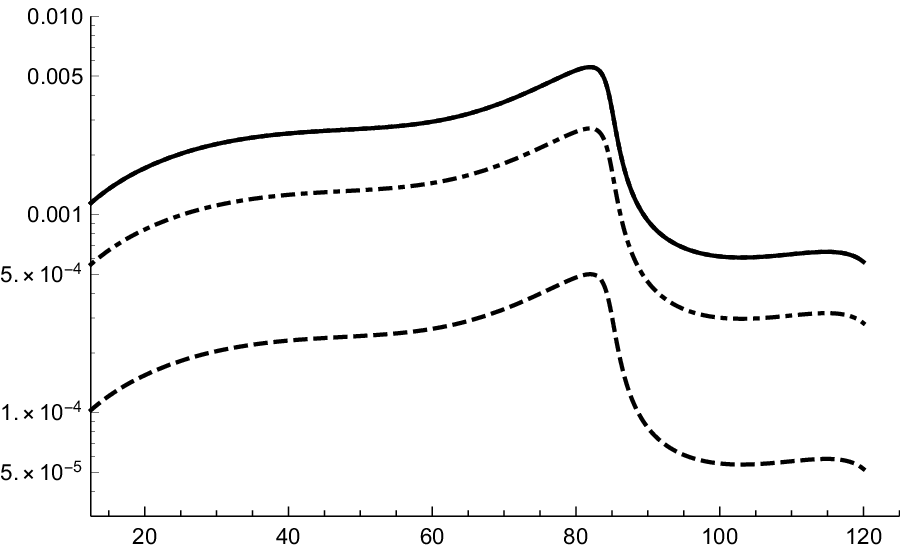}
\\
\hspace{5cm}
m_{ff}\; \text{[GeV]} &
\hspace{5cm}
m_{f\gamma}\; \text{[GeV]}
\end{array}$
\caption{\label{mff-mfg} Differential of the decay width as 
functions of $m_{ff}$ and $m_{f\gamma}$}
\end{figure}

We study the impact of the $U(1)_{B-L}$ extension 
of the SM on the differential decay widths as functions 
of $m_{ff}$ and $m_{f\gamma}$. The results are shown
in the Fig.~\ref{mff-mfg} with fixing $M_Z' =1000$ GeV. 
In these figures, the solid line
shows for the SM case by setting $c_{\alpha}=1$.
While the dashed line presents for $c_{\alpha}=0.7$
and the dashed dot line is for $c_{\alpha}=0.3$. 
In the left figure, we observe photon pole 
at the lowest region of $m_{ff}$. The decay rates decrease 
up to $m_{ff}\sim 60$ GeV. They then grown up to 
$Z$-peak (the peak of $Z\rightarrow f\bar{f}$)
which locates around $m_{ff}\sim 90$ GeV. Beyond the peak, 
the decay rates decrease rapidly. 
In the right figure, the decay widths increase 
up to peak of $m_{f\gamma} \sim 81$ GeV which is
corresponding the photon recoil mass at $Z$-pole. 
They also  decrease rapidly beyond the peak. 
It is interested to find 
that the contributions of $U(1)_{B-L}$ extension 
are sizable in both case of $c_{\alpha}=0.7, 0.3$.
These effects can be probed clearly at the
future colliders.  

\begin{figure}[]
\centering
\vspace{1cm}
\hspace{-12cm}
$\Gamma$ [KeV]\\
\includegraphics[width=12cm, height=8cm]
{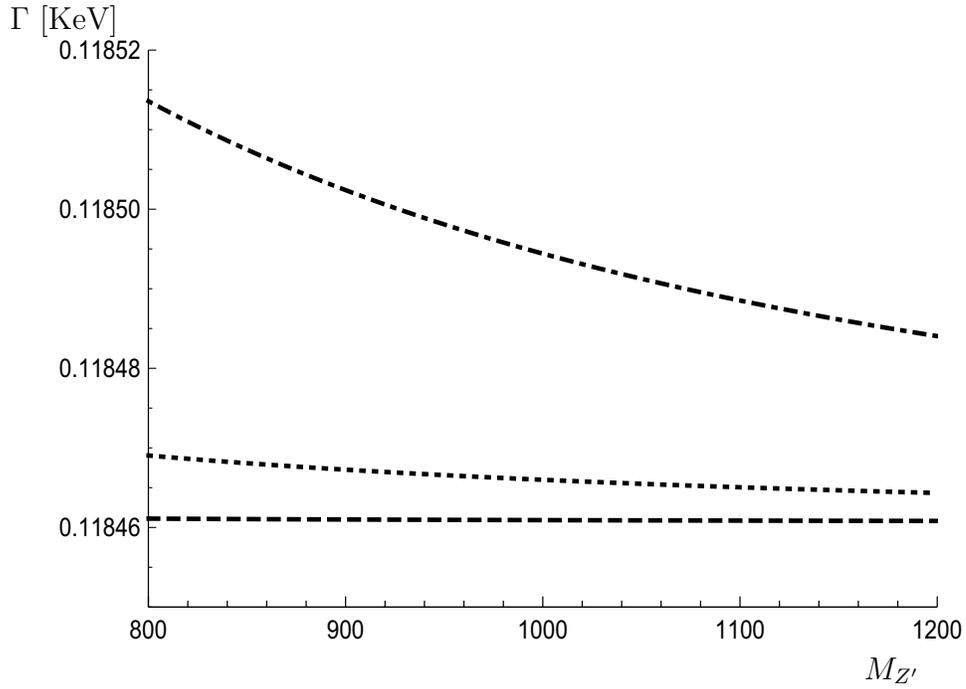}
\\
\hspace{10cm}
$M_{Z'}$
\caption{\label{gammaMZ'} The decay widths 
which are functions of $M_Z'$  }
\end{figure}

We next examine the decay widths as a function of $M_Z'$. 
In this study, we change the mass of $Z'$ boson 
as $800$ GeV $\leq M_{Z'}\leq1000$ GeV 
and set $c_{\alpha} =0.7$ (see Fig.~\ref{gammaMZ'}), 
the dashed line shows for the case of $g'_1 =0.05$, 
the dot line presents for $g'_1=0.2$ and 
the dashed dot line is for $g'_1=0.5$. 
In all range of $M_Z'$, we observe that the decay widths
are proportional to $g_1'$ (it is also confirmed later 
analyses). They also decrease with increasing $M_Z'$.  
We conclude that the contributions from the 
$U(1)_{B-L}$ extension are massive 
and can be probed clearly at future 
colliders.

\begin{figure}[]
\centering
$
\begin{array}{cc}
\hspace{-6cm}
\Gamma \text{[KeV]}
&\hspace{-6cm}
\Gamma \text{[KeV]} \\
\includegraphics[width=8cm, height=8cm]
{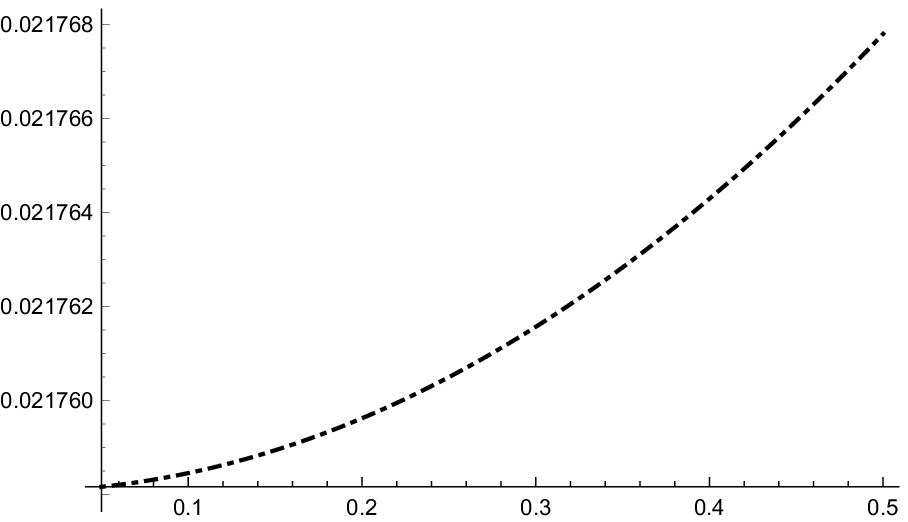}
& \includegraphics[width=8cm, height=8cm]
{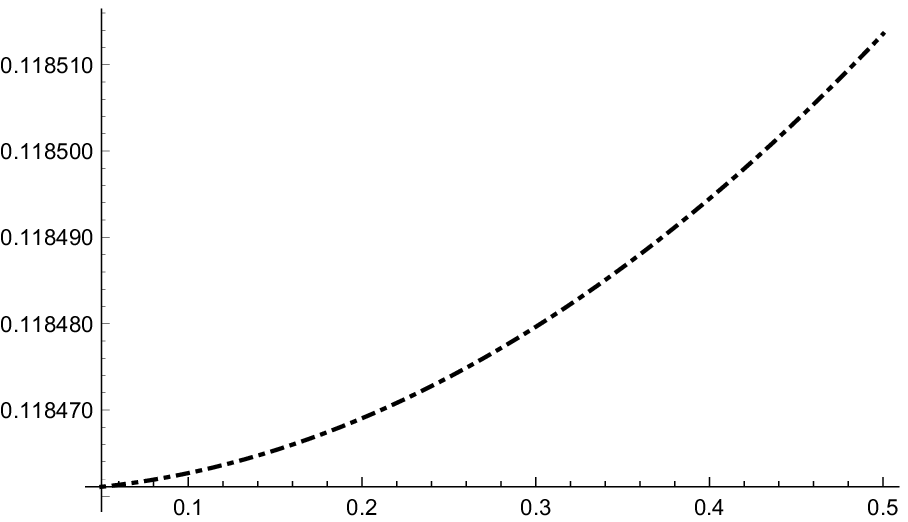}
\\
\hspace{5cm}
g_1' &
\hspace{5cm}
g_1'
\end{array}$
\caption{\label{g1} The decay widths are
function of $g_1'$}
\end{figure}

We finally discuss how the effect 
of $g'_1$ on the decay widths (seen Fig.~\ref{g1}). 
In these figures, we set $M_{Z'} =800$ GeV
and $c_{\alpha}=0.3$ ($c_{\alpha}=0.7$)
for the left figure 
(for the right figure) respectively.
We find that the decay width are 
proportional to $g_1'$. 
\subsection{Two Higgs Doublet Model}
For reviewing THDM, we refer the appendix 
$C$ for more detail. In this framework, the gauge sector
is same the SM case. It means that we have $V_i=V_j =W$,
$V_k^0 = Z, \gamma$. In the Higgs sectors, one has 
additional charged Higgs $H^+$ and two neutral 
scalar CP even Higgs bosons $H_1^0, H_2^0$, 
a CP odd Higgs $A$. All couplings relating to the 
decay processes $H\rightarrow f\bar{f}\gamma$ are 
shown in Table~\ref{thdm-coupling}.

\begin{figure}[]
\centering
$
\begin{array}{cc}
\includegraphics[width=8cm, height=8cm]
{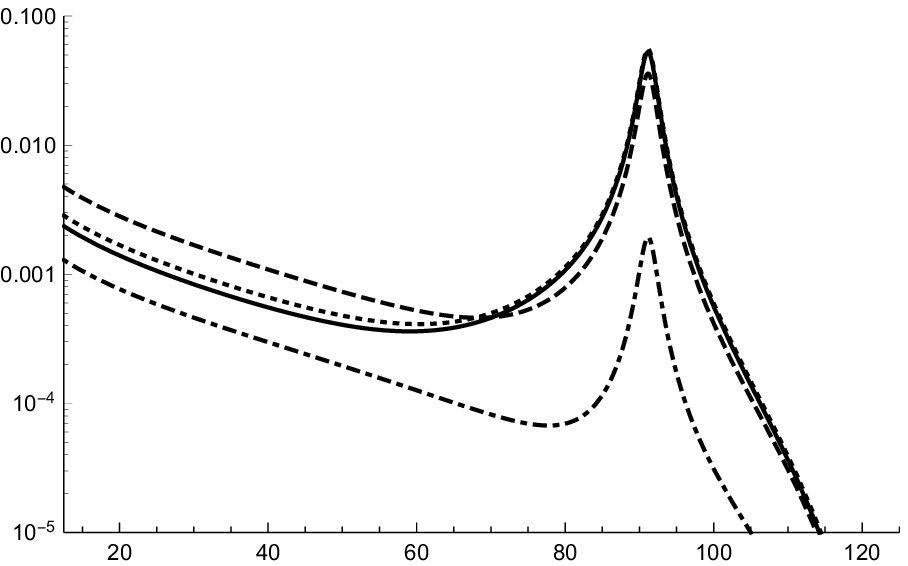}
& \includegraphics[width=8cm, height=8cm]
{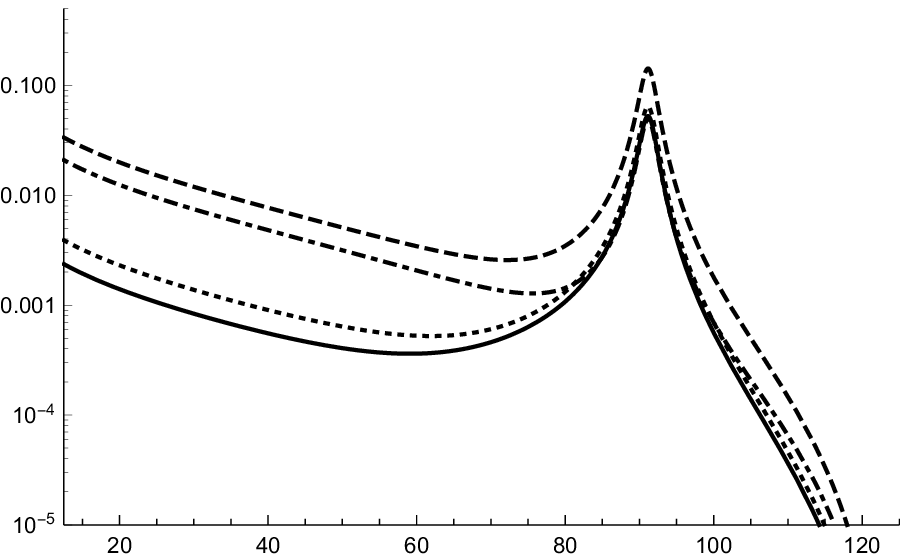}
\\  \\
\includegraphics[width=8cm, height=8cm]
{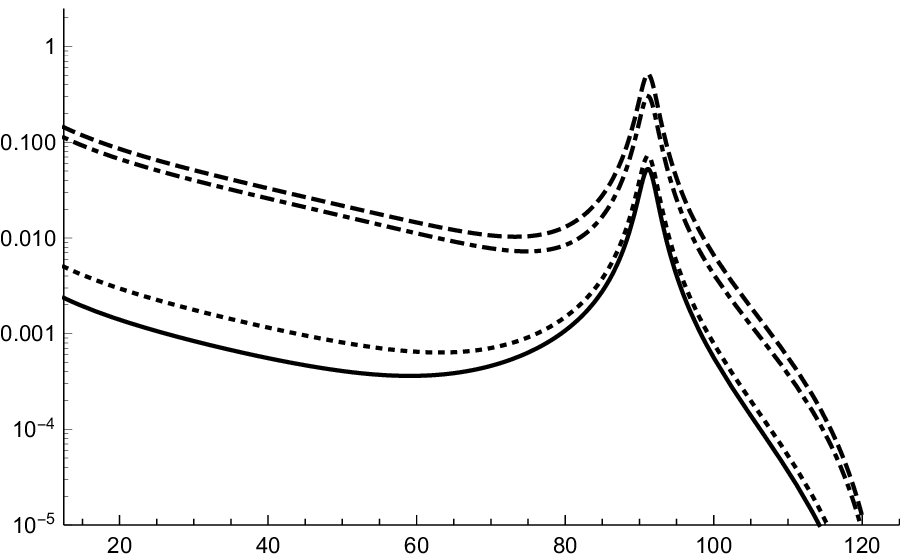}
& \includegraphics[width=8cm, height=8cm]
{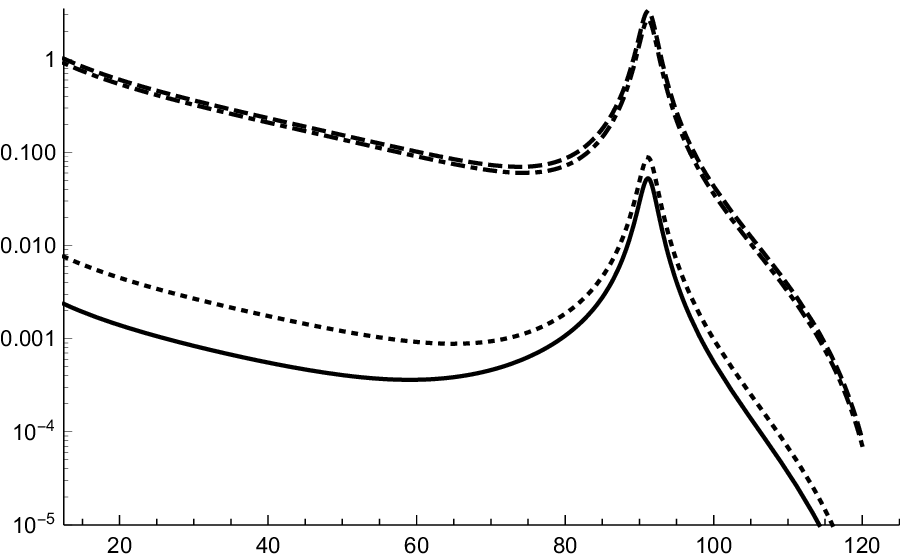}
\end{array}$
\caption{\label{mff-thdm} Differential decay rates as a 
function of the invariant mass of leptons 
$m_{ff}$ with changing $\alpha =-0.4\pi, 0, 0.4\pi$ }
\end{figure}

For the phenomenological results, we take 
$1\leq \tan\beta \leq 30$, $M_{H_{1}^0}=125.1$ 
GeV, $-900^2$ GeV$^2$ $\leq m_{12}^2 \leq 900^2$
GeV$^2$, $340$ GeV $\leq M_{H^{\pm}} \leq 900$ GeV,
$-\pi/2 \leq \alpha \leq \pi/2$. 
We study the differential decay rates as a 
function of the invariant mass of leptons 
$m_{ff}$. We select $M_{H^{\pm}} = 700$ GeV,
$ m_{12}^2 = 800^2$ GeV$^2$. The results are shown in 
Fig.~\ref{mff-thdm}. In this figure, the solid
line shows for the SM case. The dashed line presents
for the case of $\alpha = -0.4\pi$.  The dashed dot 
line is for the case of $\alpha = 0$.  
The dot line is for the case of $\alpha = 0.4\pi$. 
From the top to bottom of the figure, we have the 
corresponding figure to $\tan\beta =5, 10, 15, 25$. 
The decay rates are same behavior in previous cases,
we observe photon pole at the lowest region of $m_{ff}$. 
They decrease from photon pole to the region of 
$m_{ff}\sim 60$ GeV
and then grown up to $Z$-peak. Beyond the $Z$-peak, 
the decay rates decrease rapidly. The effect of THDM 
are also visible in all cases.    

\begin{figure}[]$
\begin{array}{cc}
\includegraphics[width=8cm, height=8cm]
{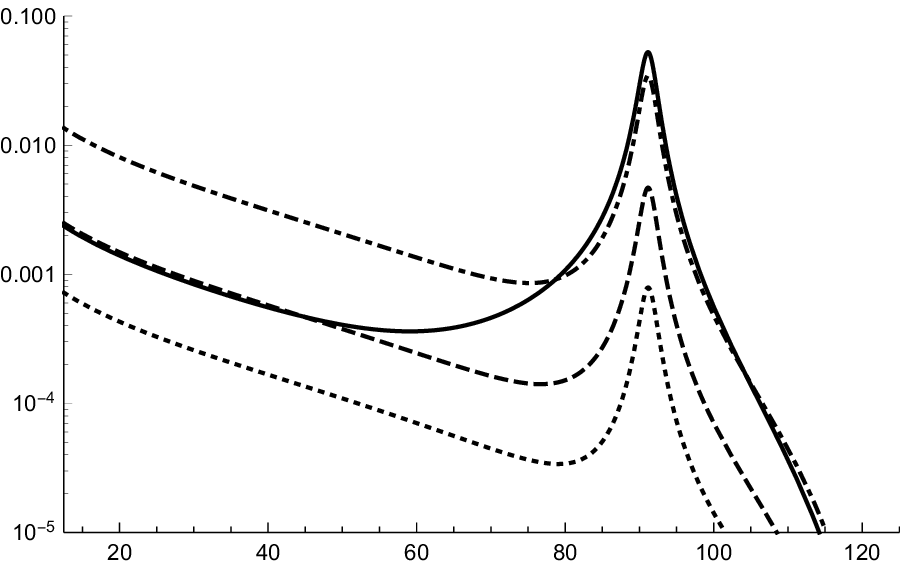}
& \includegraphics[width=8cm, height=8cm]
{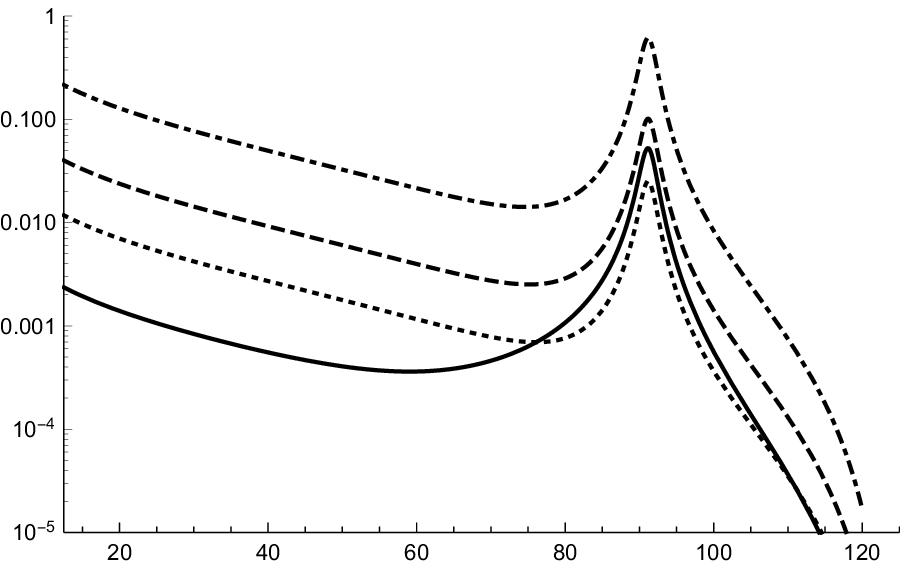}
\\  \\
\includegraphics[width=8cm, height=8cm]
{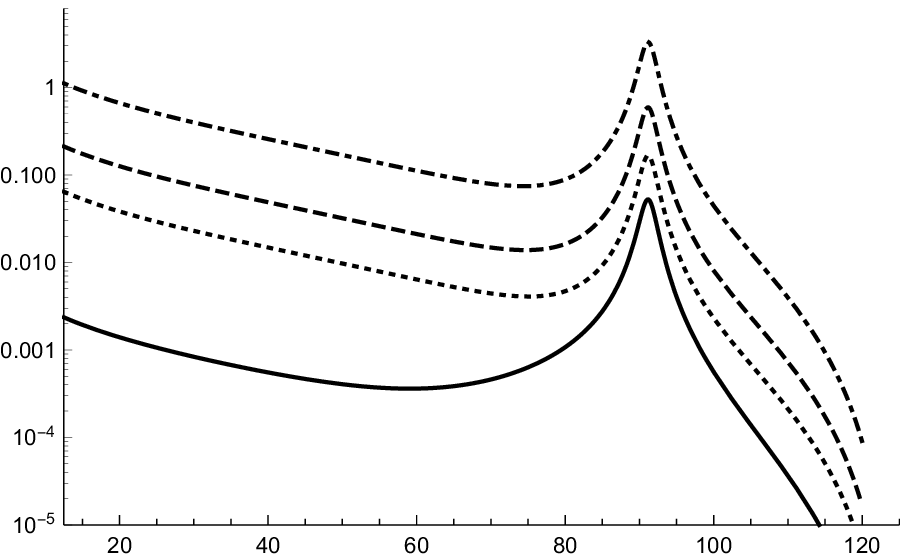}
& \includegraphics[width=8cm, height=8cm]
{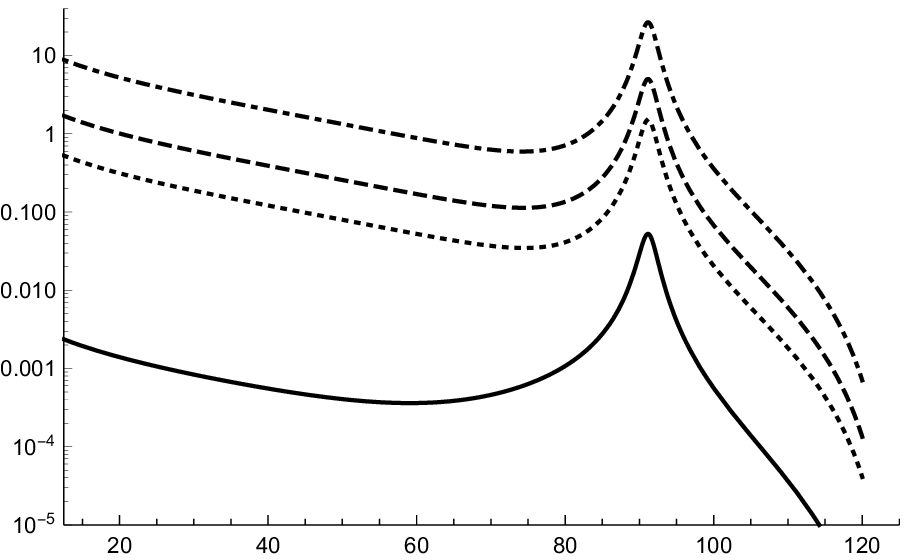}
\end{array}$
\caption{\label{MCH-thdm}Differential decay
rates as a function of the invariant mass of leptons 
$m_{ff}$ with changing charged Higgs masses
$M_{H^{\pm}} = 400, 600, 800$ GeV.}
\end{figure}

We next change charged Higgs masses 
$M_{H^{\pm}} = 400, 600, 800$ GeV and take $\alpha=0.4\pi$. 
From the top to bottom of the figure,
we have the corresponding figure to the case of 
$\tan\beta =5, 10, 15, 25$ (seen Fig.~\ref{MCH-thdm}). 
The decay widths depend on $m_{ff}$ the same previous 
explanation. They are inversely proportional to 
the charged Higgs masses.  
We find that the effect of THDM are sizeable contributions
in all the above cases. These effects can be discriminated 
clearly at future colliders. 
\section{Conclusions}   
We have performed the calculations 
for one-loop contributing to the decay 
processes $H\rightarrow f\bar{f}\gamma$
in the limit of $m_f\rightarrow0$. In this 
computation, we have considered all possible 
contributions of additional heavy vector 
gauge bosons, heavy fermions, and charged 
(and also neutral) scalar particles 
which they may exchange in the loop diagrams. 
The analytic formulas are written in terms 
of Passarino-Veltman functions which they
can be evaluated numerically by using the 
package {\tt LoopTools}. The evaluations 
have then applied standard model, 
$U(1)_{B-L}$ extension of the SM, 
two Higgs doublet model. 
Phenomenological results of the decay 
processes for the above models have 
studied in detail. We find that the 
effects of new physics are sizable 
contributions and they can be probed 
at future colliders. \\

\noindent
{\bf Acknowledgment:}~
This research is funded by Vietnam National Foundation
for Science and Technology Development (NAFOSTED) 
under the grant number $103.01$-$2019.346$. K.~H.~Phan
would like to thank Dr. L.~T.~Hue for helpful discussions.
\\
\appendix
\section{Tensor one-loop reductions}  
In this appendix, tensor one-loop 
reduction method in \cite{Denner:2005nn} 
is discussed briefly. First, the definition of
one-loop one-, two-, three- and four-point 
tensor integrals with rank $R$ are as follows:
\begin{eqnarray}
\label{tensoroneloop}
\{A; B; C; D\}^{\mu_1\mu_2\cdots \mu_R}= (\mu^2)^{2-d/2}
\int \frac{d^dk}{(2\pi)^d} 
\dfrac{k^{\mu_1}k^{\mu_2}\cdots k^{\mu_R}}{\{P_1; P_1 P_2;P_1P_2P_3; 
P_1P_2P_3P_4\}}.
\end{eqnarray}
In this formula, 
$P_j$ ($j=1,\cdots, 4$) are the 
inverse Feynman propagators
\begin{eqnarray}
 P_j = (k+ q_j)^2 -m_j^2 +i\rho.
\end{eqnarray}
In this equation, we use 
$q_j = \sum\limits_{i=1}^j p_i$ with 
$p_i$ for the external momenta, and  
$m_j$ for internal masses in the loops.
Dimensional regularization
is performed in space-time dimension 
$d=4-2\varepsilon$. The parameter $\mu^2$ 
plays role of a renormalization
scale. When the numerator of 
the integrands in Eq.~(\ref{tensoroneloop})
becomes $1$, one has the corresponding scalar
one-loop functions (they are noted 
as $A_0$, $B_0$, $C_0$ and $D_0$). 
Explicit reduction formulas for 
one-loop one-, two-, three- 
and four-point tensor integrals up to rank 
$R=3$ are written as follows~\cite{Denner:2005nn}. 
In particular, for two-point tensor integrals,
the reduction formulas are:
\begin{eqnarray}
A^{\mu}        &=& 0, \\
A^{\mu\nu}     &=& g^{\mu\nu} A_{00}, \\
A^{\mu\nu\rho} &=& 0,\\
B^{\mu}        &=& q^{\mu} B_1,\\
B^{\mu\nu}     &=& g^{\mu\nu} B_{00} + q^{\mu}q^{\nu} B_{11}, \\
B^{\mu\nu\rho} &=& \{g, q\}^{\mu\nu\rho} B_{001} 
+ q^{\mu}q^{\nu}q^{\rho} B_{111}, 
\end{eqnarray}
For three-point functions, one has
\begin{eqnarray}
C^{\mu}        &=& q_1^{\mu} C_1 + q_2^{\mu} C_2 
 = \sum\limits_{i=1}^2q_i^{\mu} C_i, 
\\
C^{\mu\nu}    &=& g^{\mu\nu} C_{00} 
 + \sum\limits_{i,j=1}^2q_i^{\mu}q_j^{\nu} C_{ij},
\\
C^{\mu\nu\rho} &=&
\sum_{i=1}^2 \{g,q_i\}^{\mu\nu\rho} C_{00i}+
\sum_{i,j,k=1}^2 q^{\mu}_i q^{\nu}_j q^{\rho}_k C_{ijk},
\end{eqnarray}
Similarly, tensor reduction formulas
for four-point functions are given by
\begin{eqnarray}
D^{\mu}        &=& q_1^{\mu} D_1 + q_2^{\mu} D_2 + q_3^{\mu}D_3 
 = \sum\limits_{i=1}^3q_i^{\mu} D_i, \\
 D^{\mu\nu}    &=& g^{\mu\nu} D_{00} 
 + \sum\limits_{i,j=1}^3q_i^{\mu}q_j^{\nu} D_{ij}, 
\\
D^{\mu\nu\rho} &=&
	\sum_{i=1}^3 \{g,q_i\}^{\mu\nu\rho} D_{00i}+
	\sum_{i,j,k=1}^3 q^{\mu}_i q^{\nu}_j q^{\rho}_k D_{ijk}.
\end{eqnarray}
The short notation~\cite{Denner:2005nn} 
$\{g, q_i\}^{\mu\nu\rho}$ is used 
as follows: $\{g, q_i\}^{\mu\nu\rho} = g^{\mu\nu} q^{\rho}_i 
+ g^{\nu\rho} q^{\mu}_i + g^{\mu\rho} q^{\nu}_i$ in the 
above relations. The scalar coefficients 
$A_{00}, B_1, \cdots, D_{333}$
in the right hand sides of the above equations 
are so-called Passarino-Veltman functions
~\cite{Denner:2005nn}. They have been 
implemented into {\tt LoopTools}~\cite{Hahn:1998yk} for 
numerical computations. 
\section{Review of $U(1)_{B-L}$ extension}
In this appendix, we review briefly $U(1)_{B-L}$ 
extension~\cite{Basso:2008iv}. This model follows 
gauge symmetry $SU(3)_C \otimes SU(2)_L \otimes 
U(1)_Y\otimes U(1)_{B-L}$. 
By including complex scalar $S$, general scalar 
potential is given by 
\begin{eqnarray}
 V(\Phi, S) &=&  m^2 \Phi^{\dag} \Phi 
 + \lambda_1 (\Phi^{\dag} \Phi)^2 +\mu^2 |S|^2 
 + \lambda_2 |S|^4 + \lambda_3 \Phi^{\dag} \Phi |S|^2.
\end{eqnarray}
In order to find mass spectrum of the scalar sector, we 
expand the scalar fields around their vacuum as follows:
\begin{eqnarray}
\Phi =
\begin{pmatrix}
\phi^+\\
\dfrac{v_{\phi} +h + i \xi}{\sqrt{2}}
\end{pmatrix}
, \quad S = \dfrac{v_{s} + h' +i \xi'}{\sqrt{2}}.
\end{eqnarray}
The Goldstone bosons $\phi^{\pm}, \xi$ will give the masses 
of $W^{\pm}$ and $Z$ bosons. In unitary gauge, the mass 
eigenvalues of neutral Higgs are given:
\begin{eqnarray}
\begin{pmatrix}
 h_1\\
 h_2
\end{pmatrix}
= 
\begin{pmatrix}
c_{\alpha}& -s_{\alpha}\\
s_{\alpha}&   c_{\alpha}\\
\end{pmatrix}
\begin{pmatrix}
 h\\
 h'
\end{pmatrix}
\end{eqnarray}
with mixing angle
\begin{eqnarray}
s_{2\alpha} = \sin2\alpha = \dfrac{\lambda_3 v_{\phi} v_{s}}
{ \sqrt{(\lambda_1 v_{\phi}^2  - \lambda_2 v_{s}^2)^2 
+ (\lambda_3 v_{\phi} v_{s})^2}  } 
\end{eqnarray}
After this transformation, masses of scalar Higgs are given
\begin{eqnarray}
 M^2_{h_1} &=& \lambda_1 v_{\phi}^2  + \lambda_2 v_{s}^2 -
 \sqrt{(\lambda_1 v_{\phi}^2  - \lambda_2 v_{s}^2)^2  +
 (\lambda_3 v_{\phi} v_{s})^2},\\ 
 M^2_{h_2} &=& \lambda_1 v_{\phi}^2  + \lambda_2 v_{s}^2 +
 \sqrt{(\lambda_1 v_{\phi}^2  - \lambda_2 v_{s}^2)^2 +
 (\lambda_3 v_{\phi} v_{s})^2}.
\end{eqnarray}
For gauge boson masses, $W, Z$ and $Z'$ 
bosons are obtained their masses
by expanding the following kinematic terms
\begin{eqnarray}
 \mathcal{L}_{\text{gauge}} \rightarrow
 (D^{\mu}\Phi)^+D_{\mu}\Phi, \quad (D^{\mu}S)^+D_{\mu}S.
\end{eqnarray}
As a result, mass of $Z'$ is $M_{Z'} = 2 v_{s}g_1'$.

The Yukawa interaction involving right-handed neutrinos
are given:
\begin{eqnarray}
\mathcal{L}_Y &=&
-y^d_{jk} \bar{q}_{Lj}d_{Rk} \Phi 
-y^u_{jk} \bar{q}_{Lj}u_{Rk} i\sigma_2\Phi^* 
-y^e_{jk} \bar{l}_{Lj}e_{Rk} \Phi 
\nonumber\\
&& -y^{\nu}_{jk} \bar{l}_{Lj}\nu_{Rk} i\sigma_2\Phi^* 
-y^{M}_{jk} \overline{(\nu_R)}_{j}^c\nu_{Rk}\; S + \text{h.c}. 
\end{eqnarray}
for $j,k=1,2,3$.
The last term in this equation is Majorana mass terms for right-handed
neutrinos. After the spontaneous breaking symmetry, the mass matrix
of neutrinos is 
\begin{eqnarray}
M = 
\begin{pmatrix}
0           & m_D = \dfrac{(y^{\nu})^*}{\sqrt{2}}v_{\phi}
\\
m_D^T       & M = \sqrt{2}y^M v_{s} 
\\ 
\end{pmatrix}
.
\end{eqnarray}
The diagonalization is obtained by the transformation
\begin{eqnarray}
\text{diag}\Big(-\dfrac{(m^i_D)^2}{M^i}, M^i \Big)= 
\begin{pmatrix}
\cos\alpha_i & -\sin\alpha_i \\
\sin\alpha_i &  \cos\alpha_i \\
\end{pmatrix}
\begin{pmatrix}
 0& m_D^i \\
 m_D^i& M^i \\
\end{pmatrix}
\begin{pmatrix}
\cos\alpha_i & \sin\alpha_i \\
-\sin\alpha_i &  \cos\alpha_i \\
\end{pmatrix}
\end{eqnarray}
for $i=1,2,3$ and $\alpha_i= \text{arcsin}(m_D^i/M^i)$. 

We show all relevant couplings in the decay under consider
In the case of $\alpha_i \rightarrow 0$, all the couplings
are shown in Table~\ref{U1-coupling}
\begin{center}
\begin{table}[h!]
\centering
{\begin{tabular}{l@{\hspace{4cm}}l }
\hline \hline
\textbf{Vertices} & \textbf{Couplings}\\
\hline \hline 
$h_{1}(h_2)f\bar{f}$ &
$-i\dfrac{c_{\alpha} (s_{\alpha})e\; m_f}{2 M_W s_W}$
\\ \\
$Z'_{\mu} f\bar{f}$  & 
$i Q_f\; g'_1\gamma_{\mu}$
\\  \\
$h_{1}(h_2) W^+_{\mu}W^-_{\nu}$ &
$i \frac{eM_W}{s_W} \;c_{\alpha} (s_{\alpha}) g_{\mu\nu}$
\\  \\ 
$h_{1}(h_2) Z_{\mu}Z_{\nu})$ 
&
$i\frac{eM_W}{c_W^2\; s_W} \;c_{\alpha} (s_{\alpha}) g_{\mu\nu}$
\\   \\
$h_{1}(h_2) Z'_{\mu}Z'_{\nu})$ &
$-i 4g'_1 M_Z' \; s_{\alpha} (-c_{\alpha})g_{\mu\nu}$ \\ 
\hline \hline
\end{tabular}}
\caption{
\label{U1-coupling}
All the couplings involving the decay processes
$H\rightarrow f\bar{f} \gamma$ in the $U(1)_{B-L}$ 
extension of the SM.}
\end{table}
\end{center}
\section{Review of Two Higgs Doublet Model}
Base on Ref.~\cite{Branco:2011iw}, we review briefly 
the two Higgs doublet model with breaking 
softly $Z_2$-symmetry. In this model, there are 
two scalar doublets $\Phi_1, \Phi_2$ with hypercharge 
$Y=1/2$. Parts of Lagrangian extended from the SM 
are presented as follows: 
\begin{eqnarray}
 \mathcal{L} = \mathcal{L}_{K} 
 + \mathcal{L}_Y - V(\Phi_1, \Phi_2).
\end{eqnarray}
Where the kinematic term is $\mathcal{L}_{K}$, 
the Yukawa part is $\mathcal{L}_Y$ 
and $V(\Phi_1, \Phi_2)$ is Higgs potential. 
First, the kinematic term is taken the form of
\begin{eqnarray}
 \mathcal{L}_{K} = \sum\limits_{k=1}^2 \left(D_{\mu}\Phi_k
 \right)^{\dag}\left(D^{\mu}\Phi_k\right)
\end{eqnarray}
with $D_{\mu} = \partial_{\mu}-ig T^a W^a_{\mu}-i\frac{g'}{2} B_{\mu}$. 
The Higgs potential with breaking the $Z_2$-symmetry 
is expressed:
\begin{eqnarray}
\label{VHiggs}
 V(\Phi_1, \Phi_2) &=& 
\frac{1}{2} 
m_{11}^2 \Phi_1^{\dag} \Phi_1  
-m_{12}^2(\Phi_1^{\dag} \Phi_2 + \Phi_2^{\dag} \Phi_1) 
+
\frac{1}{2} m_{22}^2 \Phi_2^{\dag} \Phi_2 
+ \dfrac{\lambda_1}{2} \left(\Phi_1^{\dag} \Phi_1\right)^2 
+ \dfrac{\lambda_2}{2} \left(\Phi_2^{\dag} \Phi_2\right)^2 
\n
\\
&&
\hspace{-0.2cm}
+ \lambda_3\left(\Phi_1^{\dag} \Phi_1\right) \left(\Phi_2^{\dag} \Phi_2\right) 
+ \lambda_4\left(\Phi_1^{\dag} \Phi_2\right) \left(\Phi_2^{\dag} \Phi_1\right) 
+ \frac{\lambda_5}{2}\left[\left(\Phi_1^{\dag} \Phi_2\right)^2 +
\left(\Phi_1^{\dag} \Phi_2\right)^2
\right].  
\end{eqnarray}
In this potential, $m_{12}^2$ plays role of soft broken scale of 
the $Z_2$-symmetry. The two scalar doublet fields can be 
parameterized as follows: 
\begin{eqnarray}
\Phi_1 =
\begin{pmatrix}
\phi_1^{+}\\
\dfrac{v_1 + \eta_1 + i \xi_1}{\sqrt{2}}
\end{pmatrix}
, \quad
\Phi_2 =
\begin{pmatrix}
\phi_2^{+}\\
\dfrac{v_2 + \eta_2 + i \xi_2}{\sqrt{2}}
\end{pmatrix}
.
\end{eqnarray}
We get a system equation for these parameters
from the stationary conditions of 
the Higgs potential.
The relations are shown as follows: 
\begin{eqnarray}
\label{minV}
 m_{11}^2 - \mu^2 \frac{v_2^2}{v^2} + \frac{\lambda_1}{2}v_1^2 
 + \frac{\lambda_{345}}{2}v_2^2  &=& 0, \\
 m_{22}^2 - \mu^2 \frac{v_1^2}{v^2} + \frac{\lambda_2}{2}v_2^2 
 + \frac{\lambda_{345}}{2}v_1^2  &=& 0. 
\end{eqnarray}
Where $v^2 =v_1^2+v_2^2$ is fixed at electroweak scale 
or $v= (\sqrt{2}G_F)^{-1/2}= 246$ GeV and new parameter
$\mu^2$ is defined as $\mu^2= \frac{v^2}{v_1v_2}m_{12}^2$. 
The shorten notation is $\lambda_{345}=\lambda_3+\lambda_4+\lambda_5$. 
The mixing angle is given 
$t_{\beta} = \tan\beta = v_2/v_1$. 
The mass terms of the Higgs potential $V_{mass}$ can
be expressed as:
\begin{eqnarray}
V_{mass} &=& (\phi^{+}_1, \phi^{+}_2) R_{\beta} 
\begin{pmatrix}
0 & 0\\
0 & M^2_{H^{+}}\\
\end{pmatrix}
R_{\beta}^{-1} 
\begin{pmatrix}
\phi_1^{+}\\
\phi_2^{+}
\end{pmatrix}
+\frac{1}{2}
 (\xi_1, \xi_2) R_{\beta} 
\begin{pmatrix}
0 & 0\\
0 & M^2_{A}\\
\end{pmatrix}
R_{\beta}^{-1} 
\begin{pmatrix}
\xi_1\\
\xi_2\\
\end{pmatrix}
\n \\
&&
+\frac{1}{2}
 (\eta_1, \eta_2) R_{\beta} 
\begin{pmatrix}
M^2_{H_2^0} & 0\\
0 & M^2_{H_1^0}\\
\end{pmatrix}
R_{\beta}^{-1} 
\begin{pmatrix}
\eta_1\\
\eta_2
\end{pmatrix}
.
\end{eqnarray}
Here diagonalized matrix of neutral mass 
is defined as
\begin{eqnarray}
\text{diag}(M^2_{H_2^0}, M^2_{H_1^0}) 
= R_{\alpha} \mathcal{M}^2 R_{\alpha}^T,  
\quad \text{with} \quad
(\mathcal{M}^2)_{ij} = 
\dfrac{\partial^2 V}{\partial{\eta_i} \partial{\eta_j}}. 
\end{eqnarray}
The mass eigenstates can be then expressed 
as follows:
\begin{eqnarray}
\begin{pmatrix}
G^{+}\\
H^{+}
\end{pmatrix}
= R_{\beta}^{-1} 
\begin{pmatrix}
\phi_1^{+}\\
\phi_2^{+}
\end{pmatrix}, 
\quad 
\begin{pmatrix}
G^{0}\\
A
\end{pmatrix}
= 
R_{\beta}^{-1} 
\begin{pmatrix}
\xi_1\\
\xi_2
\end{pmatrix}, 
\quad 
\begin{pmatrix}
H_2^{0}\\
H_1^{0}
\end{pmatrix}
= 
R_{\alpha}^{-1} R_{\beta}^{-1} 
\begin{pmatrix}
\eta_1\\
\eta_2
\end{pmatrix}
\end{eqnarray}
where 
\begin{eqnarray}
R_{\beta} =
\begin{pmatrix}
c_{\beta} & s_{\beta}\\
-s_{\beta} & c_{\beta}\\
\end{pmatrix}
,
\quad 
R_{\alpha} =
\begin{pmatrix}
c_{\alpha} & s_{\alpha}\\
-s_{\alpha} & c_{\alpha}\\
\end{pmatrix}
\end{eqnarray}
with $-\pi/2 \leq \alpha \leq \pi/2$.
In the unitary gauge, it is well-known that 
$G^+$ and $G^0$ are massless Goldstone bosons 
will become the longitudinal polarization of $W^+$ and $Z^0$. 
The remains $H^{\pm}$, $A$ and $H^{0}_{1,2}$ become 
the charged Higgs bosons, a CP-odd Higgs boson and 
CP-even Higgs bosons respectively. The masses of these
scalar bosons are given by
\begin{eqnarray}
 M^2_{H^{\pm}} &=& \mu^2- \frac{v^2}{2}(\lambda_{4}+\lambda_5), \\
 M^2_{A} &=& \mu^2- v^2\lambda_{5}, \\
 M^2_{H^{0}_1} &=& s_{\alpha}^2 \mathcal{M}^2_{11} 
 - 2 s_{\alpha} c_{\alpha} \mathcal{M}^2_{12} 
 + c_{\alpha}^2 \mathcal{M}^2_{22} ,\\
 M^2_{H^{0}_2} &=& c_{\alpha}^2 \mathcal{M}^2_{11} 
 +2 s_{\alpha} c_{\alpha} \mathcal{M}^2_{12} 
 + s_{\alpha}^2 \mathcal{M}^2_{22}. 
 \end{eqnarray}
From the Higgs potential in Eq.~(\ref{VHiggs})
with the stationary conditions in (\ref{minV}), 
we have $7$ parameters. They are
\begin{eqnarray}
 \Big\{ \lambda_{1,2,3,4,5}, t_{\beta}, m_{12}^2 \Big\}. 
\end{eqnarray}
For phenomenological analyses, the above parameters 
are transferred to the following parameters:
\begin{eqnarray}
 \Big\{ M^2_{H^{+}}, M^2_{A}, M^2_{H^{0}_1}, M^2_{H^{0}_2}, 
 \alpha, t_{\beta}, m_{12}^2 \Big\}. 
\end{eqnarray}
All the couplings involving the decay processes 
$H\rightarrow f\bar{f}\gamma$ 
are derived in this appendix. In general, 
we can consider the lightest Higgs boson
$H_1^{0}$ is the SM like-Higgs boson. 
In Table~\ref{thdm-coupling}, all the couplings
are shown in detail. 
\begin{center}
\begin{table}[h!]
\centering
{\begin{tabular}{l@{\hspace{2.7cm}}l }
\hline \hline
\textbf{Vertices} & \textbf{Couplings}\\
\hline \hline 
$H_1^{0} W^+_{\mu} W^-_{\nu}$
& $i \frac{2M_W^2}{v}s_{(\beta-\alpha)} g_{\mu\nu}$\\
\\
$H_1^{0} Z_{\mu} Z_{\nu}$
& $i \frac{2 M_Z^2}{v}s_{(\beta-\alpha)} g_{\mu\nu}$\\
\\
$H_1^{0}(p) H^{\pm} (q) W^{\mp}_{\mu}$
& $ \mp i \frac{M_W}{v}c_{(\beta-\alpha)}\; (p-q)_{\mu}$\\
\\
$H^{+}(p) H^{-} (q) A_{\mu}$
& $i \frac{2 M_W}{v}s_{W}\; (p-q)_{\mu}$\\
\\
$H^{+}(p) H^{-} (q)Z_{\mu}$
& $i \frac{M_Z}{v}c_{2W}\; (p-q)_{\mu}$\\
\\
$H_1^{0} H^{+}H^-$
& $\frac{i}{v}\Big[
(2\mu^2 -2M_{H^{\pm}}^2 - M^2_{H^0_1})s_{(\beta-\alpha)}
+2(\mu^2 -  M^2_{H^0_1})\text{cot}2\beta \; c_{(\beta-\alpha)}
\Big]$\\
\\
$H^{+} H^{-}A_{\mu}A_{\nu}$
& $2 i e^2\; g_{\mu\nu}$\\
\\
$H^{+} H^{-} Z_{\mu}A_{\nu}$
& $ieg\frac{c_{2W}}{c_W}\; g_{\mu\nu}$
\\
\hline \hline 
\end{tabular}}
\caption{
\label{thdm-coupling}
All the couplings involving the decay processes
$H\rightarrow f\bar{f} \gamma$ in THDM. 
}
\end{table}
\end{center}
For the Yukawa part, we refer \cite{Branco:2011iw} for more detail.
Depend on the types of THDMs, we then have the couplings of
scalar fields and fermions. 
\begin{eqnarray}
 \mathcal{L}_{H_1^0 t \bar{t} } &=& -\frac{m_t}{v}\frac{c_{\alpha}}{s_{\beta}}
 \bar{t} t H_1^0. 
\end{eqnarray}
\section{Feynman rules and couplings} 
In the below Tables, we use $P_{L/R} = (1 \mp \gamma_5)/2$,
$\Gamma^{\mu \nu \lambda} (p_1, p_2, p_3) 
= g^{\mu \nu} (p_1 - p_2)^\lambda + g^{\lambda \nu}
(p_2 - p_3)^\mu + g^{\mu \lambda} (p_3 - p_1)^\nu$ 
and $S^{\mu \nu, \alpha \beta} 
= 2 g^{\mu \nu} g^{\alpha \beta} - g^{\mu \alpha} g^{\nu \beta} 
- g^{\mu \beta} g^{\nu \alpha}$ 
and $Q_V$ denotes the electric charge of the gauge
bosons $V_i, V_j$ and $Q_S$ is charge of the 
charged Higgs bosons $S_i, S_j$. 
 Moreover, 
the factor $\xi_{V_k^{0}}$ is included for covering
all possible cases of neutral gauge boson. It
can be $0$ and $1$ for the case of the pole of
photon and Z-boson ($Z'$-boson) respectively.
\begin{center}
\begin{table}[]
\centering
{\begin{tabular}{l@{\hspace{2.7cm}}l }
\hline \hline
\textbf{Particle types} & \textbf{Propagators}\\
\hline \hline 
Fermions $f$ 
& $i\dfrac{\slashed{k} + m_{f}}{k^2-m_{f}^2}$ \\
Charged (neutral) gauge bosons $V_i (V^0_i)$
& $\dfrac{- i}{p^2 - M_{V_i}^2 (M_{V^0_i}^2)} 
\Bigg[ g^{\mu \nu} - \dfrac{p^\mu p^\nu}{M_{V_i}^2 (M_{V^0_i}^2)}  \Bigg]$ \\ 
Gauge boson $V^{0*}_k$ poles
& $\dfrac{- i}{ p^2 - M_{V_k^{0}}^2 + i\Gamma_{ V^{0}_{k} } M_{V_k^{0}} } 
\Bigg[ g^{\mu \nu} - \xi_{V_k^{0}} \dfrac{p^\mu p^\nu}{M_{V^{0}_k}^2}  \Bigg]$ \\ 
Charged (neutral) scalar bosons $S_i(S_i^0)$
& $\dfrac{i}{p^2 - M_{S_i}^2 (M_{S^0_i}^2)}$
\\  
\hline \hline \\
\end{tabular}}
\caption{\label{Feynman rules table} 
Feynman rules involving the decay channels
in the unitary gauge. 
}
\end{table}
\end{center}
\begin{table}[h!]
\begin{center}
\begin{tabular}{l@{\hspace{2cm}}l} 
\hline \hline
\textbf{Vertices} & \textbf{Couplings}   \\ \hline \hline
$H \cdot S_i  \cdot S_j $
& 
$-i \, g_{HS_iS_j} $
\\ \\
$H \cdot S^0_i  \cdot S^0_j $
& 
$-i \, g_{HS^0_iS^0_j} $
\\ \\
$H \cdot V_i^\mu \cdot V_j^\nu $  
& $i \, g_{HV_iV_j} \, g^{\mu \nu}$ 
\\ \\
$H \cdot V_i^{0\,\mu} \cdot V_j^{0\,\nu} $ 
& 
$i \, g_{HV^0_iV^0_j} \, g^{\mu \nu}$ 
\\ \\
$H \cdot \bar{f_i}  \cdot f_j $ 
& $-i \, \Big( g_{Hf_if_j}^{L}
P_L + g_{Hf_if_j}^{R} P_R \Big)$ 
\\ \\
$S^0_k \cdot \bar{f}  \cdot f $ 
& $-i \, \Big( g_{S^0_k ff}^{L}
P_L + g_{S^0_k ff}^{R} P_R \Big)$ 
\\ \\
$A^\mu \cdot f_i \cdot \bar{f_i} $
&
$i e Q_f \gamma^\mu $
\\ \\
$H (p) \cdot V_i^\mu \cdot S_j (q)$ 
& $ i \, g_{HV_iS_j} \, (p-q)^\mu$ 
\\ \\
$H (p) \cdot V_i^{0\,\mu} \cdot S^0_j (q)$
&
$ i \, g_{HV^0_iS^0_j} \, (p-q)^\mu$
\\ \\
$V^{0 \; \mu}_k \cdot V_i^\nu  \cdot S^{\pm}_j $ 
& 
$\pm i \, g_{V^0_k V_iS_j} \, g^{\mu \nu}$ 
\\ \\
$V^{0 \; \mu}_k (p_1) \cdot V_i^\nu  (p_2) \cdot V_j^\lambda  (p_3)$ 
&
$-i \, g_{V^0_k V_iV_j}  \, \Gamma_{\mu \nu \lambda} (p_1, p_2, p_3)$ 
\\ \\
$A^\mu (p_1) \cdot V_i^\nu  (p_2) \cdot V_i^\lambda  (p_3)$
& $-ieQ_V\; \Gamma^{\mu \nu \lambda} (p_1, p_2, p_3)$ 
\\ \\
$V^{0 \; \mu}_k \cdot f_i \cdot \bar{f_j}$
& 
 $i \gamma^\mu \Big( g_{V^0_k f_if_j}^{L} P_L 
 + g_{V^0_k f_if_j}^{R} P_R \Big)$
\\ \\
$V^{0 \; \mu}_k \cdot A^\nu \cdot V_i^\alpha \cdot V_j^\beta$ 
& 
$-i \, eQ_V\; g_{V^0_k AV_iV_j} \, S_{\mu \nu, \alpha \beta}$ 
\\ \\
$V^{0 \; \mu}_k \cdot A^\nu \cdot S_i \cdot S_j$
& 
$-i \, eQ_S\; g_{V^0_k AS_iS_j} \, g_{\mu \nu}$
\\ \\
$V^{0 \; \mu}_k \cdot S_i  (p) \cdot S_j  (q)$  & 
$i \, g_{V^0_k S_iS_j} \, (p-q)^\mu$ 
\\ \\
$A^\mu \cdot S_i\cdot S_i $
& $i e Q_S\, (p-q)^\mu$
\\ \\
$V^{\mu}_i \cdot f \cdot \nu_f $  &
 $i \gamma^\mu \Big( g_{V_i f \nu_f}^L P_L + g_{V_i f \nu_f}^R P_R \Big) $
\\ \\
$V^{0\,\mu}_i \cdot f \cdot f $  &
 $i \gamma^\mu \Big( g_{V^0_i f f}^L P_L + g_{V^0_i f f}^R P_R \Big) $
\\ \\
$S_i\cdot \bar{f} \cdot \nu_f $  &
 $i g_{S_i f \nu_f}^{L} P_L + i g_{S_i f \nu_f}^{R} P_R $
\\ \\
$S^{*}_i \cdot f \cdot \bar{\nu}_f $  &
 $i g_{S_i f \nu_f}^R P_L + i g_{S_i f \nu_f}^L P_R $
\\ 
\hline \hline
\end{tabular}
\end{center}
\caption{All couplings involving 
the decay processes in unitary gauge. 
\label{couplings table}}
\end{table}

\end{document}